\documentclass[twocolumn,aps,pre,superscriptaddress]{revtex4}

\usepackage{textcomp}
\usepackage{amsmath}
\usepackage{amssymb}
\usepackage{mathrsfs}
\usepackage[normalem]{ulem}
\usepackage{color}
\usepackage{empheq}
\usepackage{times}
\usepackage{multirow}
 
\usepackage{graphicx}

\begin{document}

\title{Computational hardness of spin-glass problems with tile-planted solutions}  

\author{Dilina Perera}
\email[]{dilinanp@tamu.edu}
\affiliation{Department of Physics and Astronomy, Texas A\&M University, College Station, Texas 77843-4242, USA}

\author{Firas Hamze}
\affiliation{D-Wave Systems, Inc., 3033 Beta Avenue, Burnaby, British Columbia, Canada V5G 4M9}

\author{Jack Raymond}
\affiliation{D-Wave Systems, Inc., 3033 Beta Avenue, Burnaby, British Columbia, Canada V5G 4M9}

\author{Martin Weigel}
\affiliation{Centre for Fluid and Complex Systems, Coventry University, Coventry, CV1 5FB, England}

\author{Helmut G. Katzgraber}
\affiliation{Microsoft Quantum, Microsoft, Redmond, Washington 98052, USA}
\affiliation{Department of Physics and Astronomy, Texas A\&M University, College Station, Texas 77843-4242, USA}
\affiliation{Santa Fe Institute, 1399 Hyde Park Road, Santa Fe, New Mexico 87501, USA}

\begin{abstract}

  We investigate the computational hardness of spin-glass instances on a square
  lattice, generated via a recently introduced tunable and scalable approach for
  planting solutions. The method relies on partitioning the problem graph into
  edge-disjoint subgraphs, and planting frustrated, elementary subproblems that share
  a common local ground state, which guarantees that the ground state of the entire
  problem is known \emph{a priori}. Using population annealing Monte Carlo, we compare the
  typical hardness of problem classes over a large region of the multidimensional
  tuning parameter space. Our results show that the problems have a wide range of
  tunable hardness. Moreover, we observe multiple transitions in the hardness
  phase space, which we further corroborate using simulated annealing and simulated
  quantum annealing.  By investigating thermodynamic properties of these planted
  systems, we demonstrate that the harder samples undergo magnetic ordering
  transitions which are also ultimately responsible for the observed hardness
  transitions on changing the sample composition.

\end{abstract}


\maketitle

\section{Introduction}

Optimization problems with many minima occur in a multitude of fields,
including finance, engineering, materials sciences, and machine
learning. While a range of problems such as shortest path, maximum flow
and minimum spanning tree can be elegantly solved with algorithms of a
run time growing polynomially with the system size \cite{gibbons:85},
many of the most interesting problems are known to be NP hard, and, as a
consequence, polynomial-time algorithms are very unlikely to exist.
Among discrete optimization problems many systems can be mapped onto the
Ising spin glass \cite{lucas:14,stein:13} as a central object of study
in statistical physics. Its features of frustration and random disorder
that are believed to be fundamental to the existence of a spin-glass
phase \cite{binder:86} lead to a multitude of local minima separated
by barriers --- the complex (free) energy landscape --- that are also at
the heart of hard optimization problems more generally
\cite{stein:13,barahona:82}. Besides the overlap in model systems,
statistical physics methods have also proven particularly valuable in
elucidating the structure of the solution space of optimization problems
\cite{krzakala:07} and the occurrence of phase transitions in sample
hardness as suitable control parameters are tuned \cite{hartmann:05a}.
Moreover, many heuristic techniques for optimization such as simulated
annealing (SA)~\cite{kirkpatrick:83}, parallel tempering Monte
Carlo~\cite{swendsen:86,geyer:91,marinari:92,hukushima:96,katzgraber:06a},
population annealing Monte
Carlo (PAMC)~\cite{hukushima:03,machta:10,wang:15e,wang:15}, and simulated
quantum annealing (SQA) using quantum Monte
Carlo~\cite{finnila:94,kadowaki:98,santoro:02,heim:15} have been
derived from concepts in statistical and quantum physics.

Recent years have witnessed the advent of special-purpose devices for
discrete optimization. Most noteworthy are the commercially available
analog D-Wave quantum annealing devices~\cite{johnson:11,bunyk:14},
which strive to minimize Ising Hamiltonians by exploiting quantum
tunneling and superposition. The latest machine, the D-Wave 2000Q,
allows for the optimization of instances with up to $2000$ Ising
variables, although sparse connectivity of the native graph and the
noise due to control errors~\cite{pudenz:13,pudenz:15,zhu:16} pose
limitations for practical applications.  As of now, no experimental
evidence of a quantum advantage for generic optimization applications
has been discovered, while there are some early indications of a
quantum advantage when using the machine as a physical
simulator. However, a number of recent studies have shown a speedup
for the device over selected classical algorithms for specific classes
of synthetic problems~\cite{mandra:18,albash:18}. In addition to
D-Wave devices, a number of further experimental hardware-based
solvers have been introduced, for example, the Coherent Ising machine
(optical)~\cite{wang:13b,hamerly:19}, and the complementary metal-oxide
semiconductor (CMOS) based Fujitsu
Digital Annealer~\cite{matsubara:18a,tsukamoto:17} studied recently by
Aramon {\em  et al}.~\cite{aramon:19}.

With the ensuing recent growth in interest in hardware- and
software-based Ising solvers, there is an increasing demand for hard
tunable benchmark problems for performance
comparisons~\cite{comment:complexity}.  Synthetic benchmark problems
the ground states of which are known \textit{a priori} by construction,
commonly known as samples with {\em planted solutions\/}
\cite{barthel:02}, are particularly advantageous in this regard.
Ideally, the hardness of such problems should be readily tunable, and
the construction procedure should be scalable to larger system sizes
with reasonable computational effort.  A number of solution planting
schemes have been proposed for the case of short-range Ising
spin glasses. For example, Ref.~\cite{hen:15a} presented an
approach in which constraint satisfaction problems in the form of
cycles of couplers (i.e., frustrated loops) are embedded into an
arbitrary graph.  Although the problem hardness is tunable, the
range of coupler strength is uncontrolled and increases with the
system size, which leads to amplified control errors in analog devices
with finite precision such as the D-Wave machines. Reference
\cite{king:15} introduced a variation of the method with limited range
of coupling strength that mitigates the precision
limitations. However, both methods produce problems that are easier
than other standard benchmarks for a wide range of software and
hardware-based solvers~\cite{coffrin:19}.

Reference \cite{marshall:16} proposed an adaptive algorithm for
generating hard problems by iteratively updating the coupler values
based on the time to solution (TTS) of a chosen solver. The computational
resources required by the procedure, however, severely restrict the
feasible size of the problems. Moreover, the method relies on the
assumption that the problem hardness as predicted by the chosen solver
may also extend to other solvers. In the ``patch planting'' method
introduced in Ref.~\cite{wang:17}, the problem graphs are constructed by
stitching together smaller subgraphs via couplers that ensure
consistency among boundary spins.  Although the method allows for the
construction of arbitrarily large problems with no restrictions on
coupler values or the graph topology, the resultant problems are less
difficult than entirely random problems of comparable size. In addition,
as the ground states of the constituent subgraphs need to be
determined in advance, the problem construction may require considerable
computational effort depending on the size and connectivity of the
subgraphs.

Recently, some of the present authors proposed a planted solutions
scheme~\cite{hamze:18} that circumvents several weaknesses of the
aforementioned methods. Henceforth, let us refer to this approach as
``tile planting.''  The method is based on decomposing the underlying
problem graph into edge-disjoint subgraphs, and embedding a selected set
of subproblems in the resulting subgraphs.  The subproblems are chosen
such that they share a common (local) ground state, which ensures that
the ground state of the entire problem is known \textit{a priori}.  Due
to the myriad ways in which subproblems can be chosen, one can construct
problems with a wide range of computational complexities. The problem
construction demands very low computational effort, and facilitates the
generation of arbitrarily large problems on the underlying topology.

For completeness, we mention two alternate planting approaches
\cite{hamze:19x,hen:19a} that start from a set of linear equations and
then use these to plant a solution for a quadratic Ising system. In
particular, the Wishart planted ensemble \cite{hamze:19x} is amenable to
theoretical exploration, thus allowing one to make analytical
predictions for the typical computational hardness of the planted
problems with a given bit precision. While this is a huge bonus, the
drawback is that the planted ensemble requires a complete graph and is
thus not best suited to benchmark current quantum annealing hardware
where an all-to-all connectivity is extremely hard to build
\cite{bunyk:14}. However, the ensemble is ideal for the benchmarking of
(classical) digital annealer hardware \cite{matsubara:18a,tsukamoto:17},
as well as optical systems \cite{wang:13b,hamerly:19}.

In the present paper, we explore the computational hardness of planted
problems on square lattices constructed using tile planting. By focusing
on the square lattice we are able to exhaustively investigate a large
region of the tuning parameter phase space, a task that is impractical
for cubic lattices due to the comparatively larger phase space. Although
the planarity of the underlying graph renders them exactly solvable in
polynomial time using planar graph
solvers~\cite{schraudolph:09,galluccio:01,galluccio:00,saul:94}, the
free-energy landscape generally remains challenging for the majority of
practical heuristic optimization methods that are agnostic to this
special graph structure, and the resulting class of problems has many
common features with nonplanar variants of the tile-planting approach
\cite{hamze:18} in, e.g., three space dimensions or nonplanar graphs. It
should also be noted that one can break planar methods by superficial
modifications such as adding fields, but the purpose here is not to
obfuscate our results through such changes.

The remainder of this paper is organized as follows. In
Sec.~\ref{sec:method} we outline our planted solutions scheme for square
lattices, and discuss the relationship of the constructed problems to
constraint satisfaction problems.  In Sec.~\ref{sec:results} we
investigate the variations in problem difficulty across different
instance classes via the population annealing Monte Carlo method.
Particular emphasis is given to two specific regions in the parameter
space where hardness transitions are observed, which we further
investigate using simulated annealing and simulated quantum annealing.
Furthermore, we investigate the thermodynamic properties of a selected
set of instance classes and discuss the connection between hardness
and phase behavior.  Section~\ref{sec:summary} presents our concluding
remarks.

\section{Tile planting}
\label{sec:method}

We start by describing our planting scheme for square lattices, which
will be the focus of this paper. For a demonstration of the method for
cubic lattices see Ref.~\cite{hamze:18}. Note that the approach is
adaptable to arbitrary graphs, with a carefully chosen set of subproblem
classes suitable for the underlying topology.

\begin{figure}[tb!] 
  \includegraphics[width=0.5\linewidth]{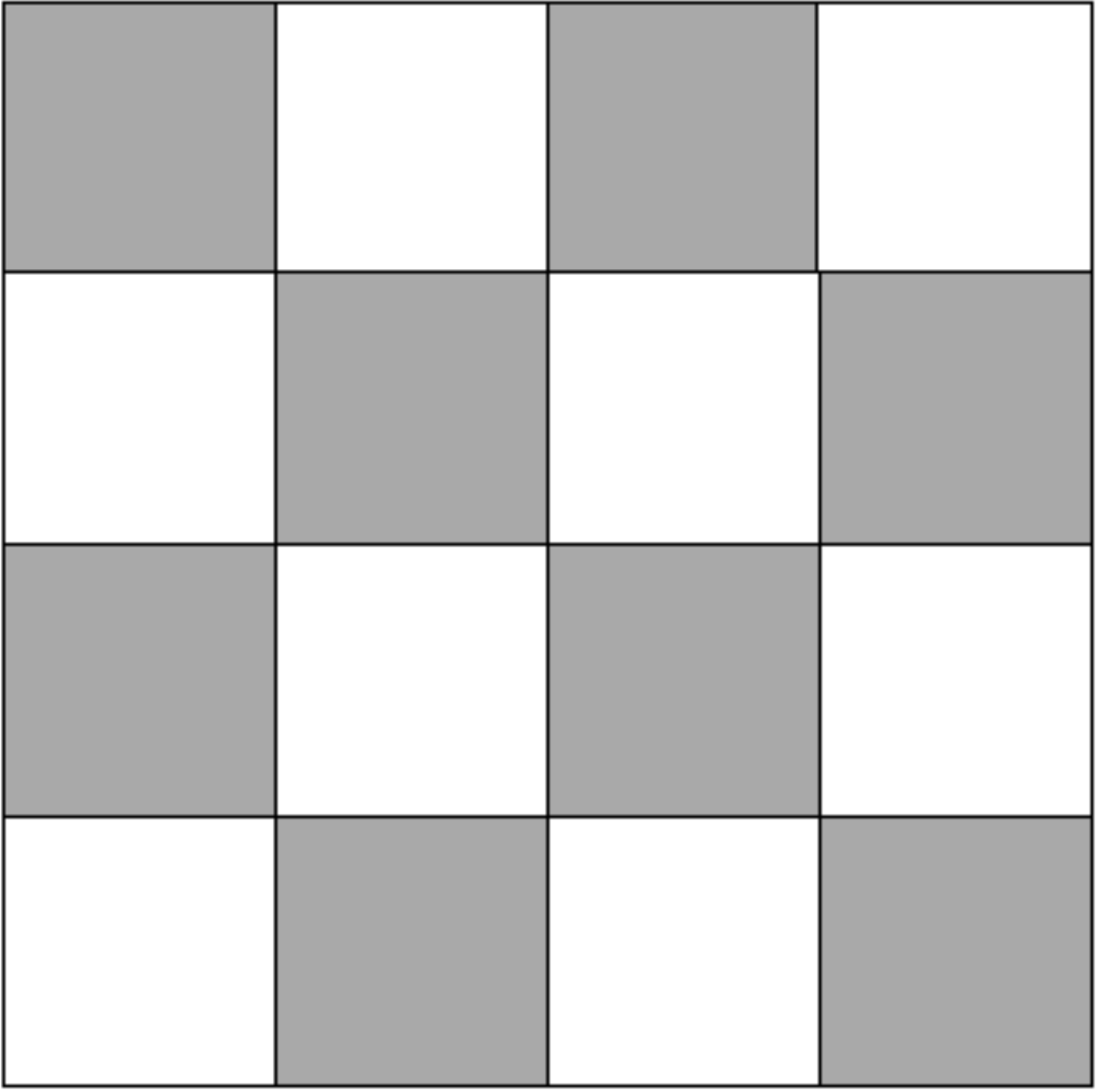}
  \caption{Decomposition of the square-lattice graph into vertex-sharing, unit-cell
    subgraphs (shaded cells). Note that subgraphs do not share any edges. When periodic
    boundary conditions are imposed, each vertex is shared by exactly two subgraphs.
    Unit-cell Ising subproblems with a common local ground state are planted on the
    subgraphs. The entire problem has a ground state (i.e., the planted
    solution) composed of the same local ground state occupying all subgraphs.}
 \label{fig:checker_board}
\end{figure}

Let us consider a square lattice of linear size $L$ with periodic
boundary conditions. We denote the underlying graph as $G = (V,E)$,
with vertex set $V$ and edge set $E$.  As illustrated in
Fig.~\ref{fig:checker_board}, by partitioning the lattice into a
checkerboard pattern, we decompose the graph $G$ into a set of
vertex-sharing, unit-cell subgraphs (represented by shaded cells).  Note
that the subgraphs do not share any edges, and with periodic boundary
conditions, each vertex in $V$ appears exactly in two subgraphs.  Let
$C$ denote the vertices of a given unit-cell subgraph, and let $\mathcal{C}$
represent the set of all such subgraph vertex sets constituting $G$.
For each subgraph $C$, we define the Ising Hamiltonian
\begin{equation} \label{eq:subropblem_hamiltonian}
	\mathcal{H}_C = -\sum_{(ij) \in E[C]} J_{ij} s_i s_ j  \qquad s_i \in \{\pm 1\},
\end{equation}
which will henceforth be referred to as a subproblem Hamiltonian.  We
then form the complete Ising Hamiltonian $\mathcal{H}$ over graph $G$ by
adding all subproblem Hamiltonians as
\begin{equation} \label{eq:total_hamiltonian}
	\mathcal{H} = \sum_{C \in \mathcal{C}} \mathcal{H}_C.
\end{equation}  
As a straightforward result of constructing $\mathcal{H}$ as a direct
sum of subproblem Hamiltonians, if all subproblem Hamiltonians share a
common ground-state configuration, this state will also be a ground
state of the overall Hamiltonian $\mathcal{H}$. Thus, the ground state
energy of $\mathcal{H}$ is also known \textit{a priori}. By carefully
choosing the couplers for the subproblem Hamiltonians from different
classes with varying levels of frustration, one can control the
computational hardness of the entire problem to a great
extent~\cite{hamze:18}.

\begin{figure}[tb!] 
  \hspace*{-5.0em}\includegraphics[width=0.5\columnwidth]{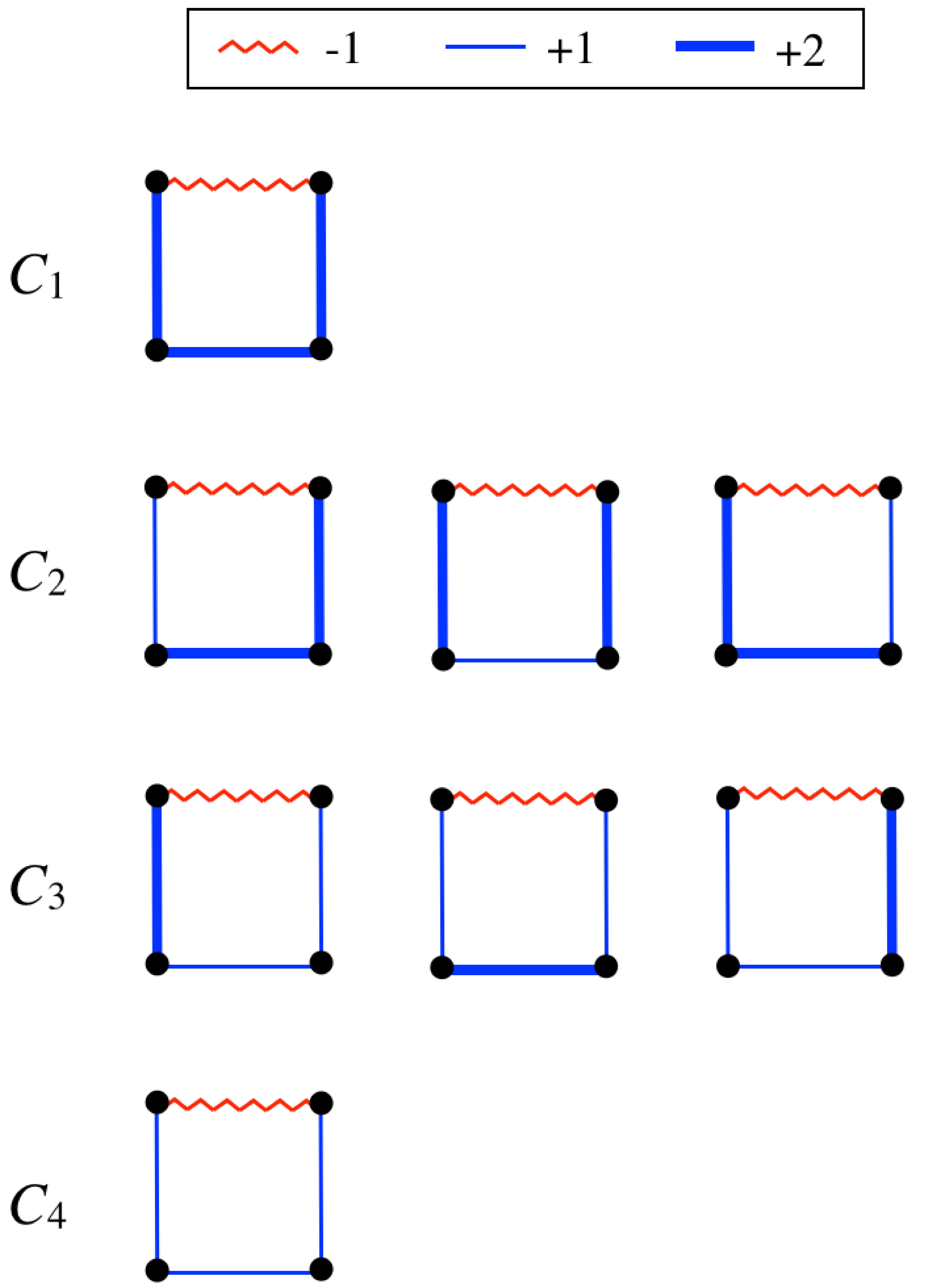}
  \caption{Class representatives of the four unit-cell subproblem types for square
    lattices. Red wiggly lines represent antiferromagnetic couplers with
    value $-1$, whereas straight blue lines represent ferromagnetic
    couplers with values $+1$ (thin lines) and $+2$ (thick lines). Black
    dots represent the spins. Subproblems of class $C_i$ have $i$ ground
    states, which always include the all-spin-up and all-spin-down
    states.}
 \label{fig:classes}
\end{figure}

We now proceed to define a set of subproblem classes for the square
lattice topology. Let us choose the ferromagnetic ground state, i.e.,
all spins either up or down as the intended shared ground state of the
subproblems. Note that once the problem is constructed one can
transform the planted solution from the ferromagnetic ground state to an
arbitrary state via a simple gauge transformation. Here we propose four
subproblem classes denoted by $\{C_i\}$, $i \in \{1, 2, 3, 4\}$,
corresponding to bond configurations on square plaquettes (see the
illustration in Fig.~\ref{fig:classes}). For the exchange couplings we
use two magnitudes $J_l$ and $J_s$ with $|J_l|>|J_s|$.  Here we
arbitrarily choose $|J_l| = 2$ and $|J_s| = 1$.  The use of two
different coupling strengths allows us to tune the degeneracy of the
overall optimization problems and hence achieve a variation in hardness
of the instances created.

An instance of the subproblem class $C_i$ is constructed as follows.
First, a chosen edge of the subproblem unit cell is set to the
antiferromagnetic coupler value of $-J_s$.  Then we assign the
ferromagnetic value $+J_s$ to a randomly selected set of $i-1$ remaining
edges, while the rest of the edges are set to $+J_l$.  This construction
procedure leads to three equivalent subproblem representations for the
classes $C_2$ and $C_3$ and unique representations for $C_1$ and $C_4$,
up to overall rotations in multiples of $90^{\circ}$ of the plaquettes
(see Fig.~\ref{fig:classes}). From this construction it is clear that
subproblems of class $C_i$ have $i$ different ground states (apart from
overall spin reversal), and the ferromagnetic configurations are always
ground states.

When generating planted instances, we first choose a subproblem class
for each unit-cell subproblem (i.e., shaded square plaquettes in
Fig.~\ref{fig:checker_board}) based on a chosen probability distribution
over classes $\{C_i\}$.  Then for each subproblem, we randomly choose one
of the equivalent class representatives shown in Fig.~\ref{fig:classes},
followed by a random rotation of the plaquette to introduce more
disorder. We define each instance class based on the chosen probability
distribution over the subproblem classes $\{C_i\}$.  Let $p_i$ denote
the probability of choosing subproblems from class $C_i$.  We can then
uniquely identify each instance class in terms of three probability
parameters, for instance, $\{p_1, p_2, p_3\}$, where $p_1+ p_2 + p_3 \leq
1$.  As we demonstrate below, one can induce tremendous variations in
problem hardness by tuning the probability parameters $\{p_i\}$.
Moreover, due to the simplistic nature of the construction procedure,
hundreds of problem instances with moderate system sizes can be
generated in seconds with very low computational cost.

\begin{figure}[tb!]
  \includegraphics[width=1.0\columnwidth]{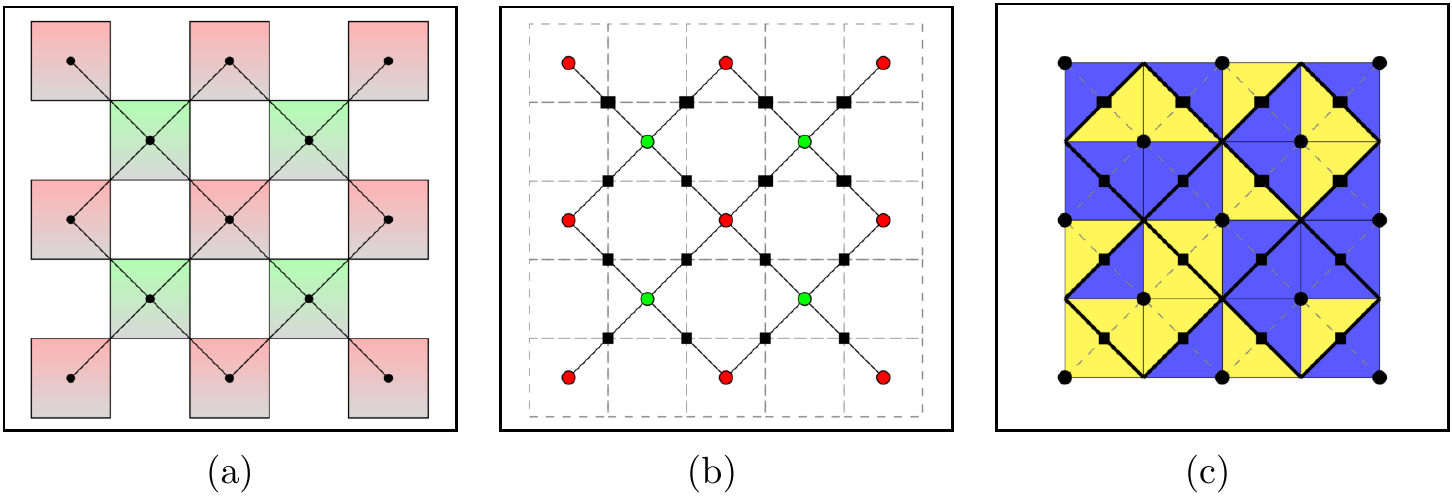}
  \centering
  \caption{ Steps involved in mapping a planted Ising problem into a tiling puzzle.
    For clarity, free boundary conditions are used. (a) Construction of the
    subproblem lattice $\tilde{G}$ by placing a vertex at the center of each planted
    cell. The domain of each subproblem lattice variable $\mathbf{s}_C$ is the set
    of ground states of the planted subproblem.  (b) The factor graph of $\tilde{G}$,
    with the factors shown as squares.  As discussed in the text, the factors
    represent the constraints enforced on the subproblem lattice variables due to
    each Ising spin being shared by two adjacent cells.  (c) The Voronoi tessellation
    of $\tilde{G}$ which yields the set of tiling locations (areas within thick
    lines). A hypothetical tile is placed at each tiling location.  The tile faces
    are in two colors; blue (dark) and yellow (light); which represent the Ising
    spin values $+1$ and $-1$. An optimal solution to the overall problem requires
    the colors of adjacent tile faces to agree (i.e., the factor constraints to be
    satisfied). Adapted from Ref.~\cite{hamze:18}.}
 \label{fig:tiling_puzzle}
\end{figure}

We note that all of the subproblems $C_i$ correspond to frustrated
squares \cite{toulouse:77a}, whereas the remaining plaquettes (white
cells in Fig.~\ref{fig:checker_board}) are randomly frustrated with
probability $15/32$, leading to an overall concentration of about 73\%
frustrated plaquettes in the resulting instances \cite{comment:gauge}.
To study the degeneracy of the full samples resulting from subproblem
classes $C_i$, we use exact enumeration based on the evaluation of
Pfaffians which can be achieved in polynomial time for the case of
planar graphs \cite{galluccio:00}. On its own, subproblem $C_i$ has $i$
ground states and hence $i^{\cal N}$ is an upper bound to the number of
ground states of samples composed of subproblems of class $i$, where
${\cal N}$ is the number of plaquettes (shaded cells in
Fig.~\ref{fig:checker_board}). As a consequence, instances entirely
composed of $C_1$ have a unique ferromagnetic ground state up to overall
spin reversal. For the other classes the connection of subproblems
provides additional constraints reducing the degeneracy. For example,
for $L=24$ we find the following ground-state entropies $s_0 = S_0/{\cal
N}$ per shaded plaquette
\begin{equation}
  \exp(s_0) = \left\{
    \begin{array}{ll}
      1.33(5) & \mbox{for}\;C_2\\
      2.285(2) & \mbox{for}\;C_3\\
      4.018(2) & \mbox{for}\;C_4 \; .\\
    \end{array}
  \right.
  \label{eq:degeneracies}
\end{equation}
In comparison, the standard $\pm J$ spin-glass model in two space
dimensions has $\exp(s_0) \approx 1.38$, so it turns out to be closest
to $C_2$ in this respect, which as we will see later on provides the
hardest samples among the pure classes $C_i$.

As demonstrated in Ref.~\cite{hamze:18}, the planted problem
construction can be translated to a specific type of constraint
satisfaction problem called a tiling puzzle~\cite{demaine:07}, a
well-known example of which is Eternity II.
Figure~\ref{fig:tiling_puzzle} illustrates the steps involved in the
derivation of a tiling puzzle from the unit-cell planting formulation.
As shown in Fig.~\ref{fig:tiling_puzzle}(a), we first construct a graph
with respect to $\mathcal{C}$ (which we call the subproblem lattice
$\tilde{G}$) by placing a vertex at the center of each planted cell.
Each pair of vertices in $\tilde{G}$ is considered adjacent if and only
if the corresponding subgraphs (shaded cells) share a corner vertex.
Let us associate each vertex in $\tilde{G}$ with a subproblem lattice
variable $\mathbf{s}_C$ over the four Ising variables of the
corresponding subproblem.  The domain of $\mathbf{s}_C$ is the set of
ground states of the subproblem.  However, viable solutions to
$\mathbf{s}_C$ are constrained by the equality conditions enforced on
the shared Ising variables.  That is, for two neighboring cells $C$ and
$C^\prime$ with a shared corner vertex $i$, $s_C^i = s_{C^\prime}^i$.  A
solution to the overall problem should simultaneously satisfy all such
equality constraints. By interpreting each equality constraint as a
factor over the subproblem lattice variables $\mathbf{s}_C$ and
$\mathbf{s}_{C^\prime}$, one can construct a factor
graph~\cite{kschischang:01} associated with $\tilde{G}$ as shown in
Fig.~\ref{fig:tiling_puzzle}(b), where the factors are represented by
squares.  We then perform a Voronoi tessellation~\cite{okabe:00} of
$\tilde{G}$ [Fig.~\ref{fig:tiling_puzzle}(c)] which partitions the
surrounding space into a set of tilted square regions centered at each
vertex, the boundaries of which are shown in thick lines.  We can interpret
each region as a tiling location, with an allowable set of tiles and
orientations as determined by the domain of $\mathbf{s}_C$ (the ground-state 
set of the subproblem).  In Fig.~\ref{fig:tiling_puzzle}(c), a
hypothetical tile is placed at each location, where blue (dark) and
yellow (light) tile faces represent the two Ising spin values ($+1$ and
$-1$). The optimal solution to the overall problem can now be
interpreted as finding a tile arrangement in which no two adjacent tile
faces have a different color; i.e., all factor constraints are satisfied.

\section{Results} 
\label{sec:results}

\subsection{Simulation details}

To compare the typical hardness of planted instances across different
problem classes, we primarily use population annealing 
Monte Carlo~\cite{hukushima:03,machta:10,wang:15e,wang:15,barash:17a,barzegar:18}.
For certain regimes with interesting variations in hardness, we also use
SA~\cite{kirkpatrick:83} and SQA~\cite{finnila:94,kadowaki:98,santoro:02,heim:15}.

PAMC is a sequential Monte Carlo
algorithm somewhat similar in spirit to SA. In
contrast to the latter, however, PAMC was designed with equilibrium
simulations in mind, and not as an optimization method. It considers an
ensemble of copies of the same system and disorder realization that is
started at a high temperature and successively cooled down to the lowest
temperatures of interest. In addition to equilibrating update steps
applied to each copy that are typically implemented as Markov chain
Monte Carlo, the population is further relaxed on lowering the
temperature by a resampling step that selects and replicates (or
eliminates) copies proportional to their relative Boltzmann weight at
the new temperature. Large fluctuations in the weights that determine
resampling are a signature of the difficulty in equilibrating a given
system \cite{machta:10} and lead to the descendants of only a few
original copies dominating the population. Hence, a measure of the
effective number of surviving replica families at the lowest temperature
allows one to distinguish between hard and easy problems within the
instance classes. If $n_i$ is the fraction of the population descended
from replica $i$ in the initial population (i.e., the size of the $i$th
family), the entropic family size $\rho_s$~\cite{wang:15e} is given by
\begin{equation} \label{eq:rho_s}
	\rho_s = \lim_{R \to \infty} R \: e^{\sum_i n_i \log n_i},
\end{equation} 
where $R$ is the initial population size.  For a given set of simulation
parameters, the larger the value of $\rho_s$, the smaller the number of
surviving families, and hence the more rugged the problem's energy
landscape. Thus, $\rho_s$ provides a measure of hardness for algorithms
that are based on local search in the classical energy landscape. As was
shown previously, $\rho_s$ is highly correlated with other well
established hardness metrics such as the integrated autocorrelation time
in Markov-chain Monte Carlo and, specifically, in parallel tempering
(PT) Monte Carlo \cite{wang:15e}. A unique advantage of $\rho_s$ is that
it can be measured with relatively low computational cost in comparison
to PT autocorrelation times, as well as metrics based on optimized
time to solution \cite{comment:lb}.  This allows us to rigorously
investigate a large region of the instance class parameter space that
results from random selection of subproblem types.

\begin{table}
\caption{Parameters for the population annealing Monte Carlo
    simulations. $L$ is the linear system size, $R$ is the number of
    replicas, $T_0$ is the lowest temperature, and $N_T$ is the number
    of temperature steps. At each temperature step, $N_S=10$ Monte Carlo
    sweeps per temperature step are performed on each replica. The
    annealing schedule is linear in the inverse
    temperature.\label{tab:sim_param}}
\begin{tabular*}{\columnwidth}{@{\extracolsep{\fill}} l c c r }
\hline \hline 
$L$ & $R$ & $T_0$ & $N_T$  \\ [0.5ex] 
\hline
16 & $1 \times 10^5$ & 0.2 & 301 \\
24 & $2 \times 10^5$ & 0.2 & 301 \\
32 & $1 \times 10^6$ & 0.2 & 301 \\  [1ex]
\hline
\hline
\end{tabular*}
\end{table}

Table~\ref{tab:sim_param} shows the PAMC simulation parameters for the
three linear system sizes used in our paper to characterize hardness,
$L=16$, $24$, and $32$. $\rho_s$ is known to converge with
increasing populating size $R$, and $R$ values for different system
sizes were chosen accordingly to ensure that this convergence criterion
is met by all simulations.

In addition to PAMC simulations, we also perform SA and SQA calculations
on a selected subset of problem classes.  SA calculations are performed
with the optimized multispin code by Isakov \textit{et
al.}~\cite{isakov:15}, using a linear annealing schedule with an initial
temperature of $T_\mathrm{i} = 10$ and a final temperature of
$T_\mathrm{f}=0.2$. For the SQA calculations, we use $64$ Trotter slices
and a temperature of $T=0.03125$, while linearly decreasing the
transverse field magnitude from $\Gamma=2.5$ to $0$.  For both SA and
SQA we estimate the time required to find the optimal solution at least
once with probability $p_d$, i.e., the
time to solution~\cite{ronnow:14a,boixo:14}:
\begin{equation}
\text{TTS}(t_\text{A}) = t_\text{A} \frac{\log(1-p_d)}{\log[1-p_s(t_\text{A})]},
\end{equation}
where $t_\text{A}$ is the annealing time and $p_s$ is the success
probability, while $p_d = 0.99$.  We measure the TTS for individual
problem instances, and then calculate the median of the distribution of
TTS values for a given instance class and system size.  We minimize the
median TTS in annealing time $t_\text{A}$ to obtain the optimal median
TTS which takes into account the best tradeoff between the annealing
time and the number of independent runs~\cite{ronnow:14a}.  For
simplicity, hereafter we use the term time to solution to refer to
the aforementioned {\em optimal median} TTS obtained for each instance
class and system size.

Finally, to gain insight into the physical mechanisms of hardness, we
investigate thermodynamic properties of a selected subset of instance
classes using PAMC, as well as using the Pfaffian enumeration method of
Ref.~\cite{galluccio:00} to determine the exact density of states and
hence exact thermodynamic averages for energetic quantities for each
sample. Via PAMC simulations, we measure the Binder ratio
$g_{\mathcal O}$ via
\begin{equation}
g_\mathcal{O} = \frac{1}{2}\left(3 - [\langle \mathcal{O}^4 \rangle]_\text{av} /
[\langle \mathcal{O}^2 \rangle]^2_\text{av}\right),
\end{equation}
where the order parameter $\mathcal{O}$ represents either the
magnetization $m = (1/N)\sum_i s_i$ (ferromagnetic order), or the spin
overlap $q=(1/N)\sum_i s_i^\alpha s_i^\beta$ (spin-glass order), with
$\alpha$ and $\beta$ representing two independent replicas of the system
with the same disorder. Note that $\langle \cdots \rangle$ and $[
\cdots ]_\text{av}$ represent thermal average and disorder average,
respectively. As a dimensionless quantity, the Binder ratios
for different system sizes are expected to cross close
to the point of a continuous phase transition. From the expected scaling
behavior \cite{privman:90},
\begin{equation}
  g_\mathcal{O} \sim G_\mathcal{O} [L^{1/\nu_\mathcal{O}} (T-T_c)],
\end{equation}
it is straightforward to extract the critical temperature $T_c$ and the
critical exponent $\nu_\mathcal{O}$ from a scaling collapse
\cite{houdayer:04,katzgraber:06}. In addition to the Binder ratio, we
also investigate the specific heat,
\begin{equation}
  C_v = 1/(k_B T^2 N)[ \langle \mathcal{H}^2 \rangle -\langle \mathcal{H} \rangle^2
  ]_\text{av},
\end{equation}
using the planar graph solver, which provides us with a measure free of
statistical errors.

\subsection{Base instance classes}

\begin{figure}[h]
  \includegraphics[width=\linewidth]{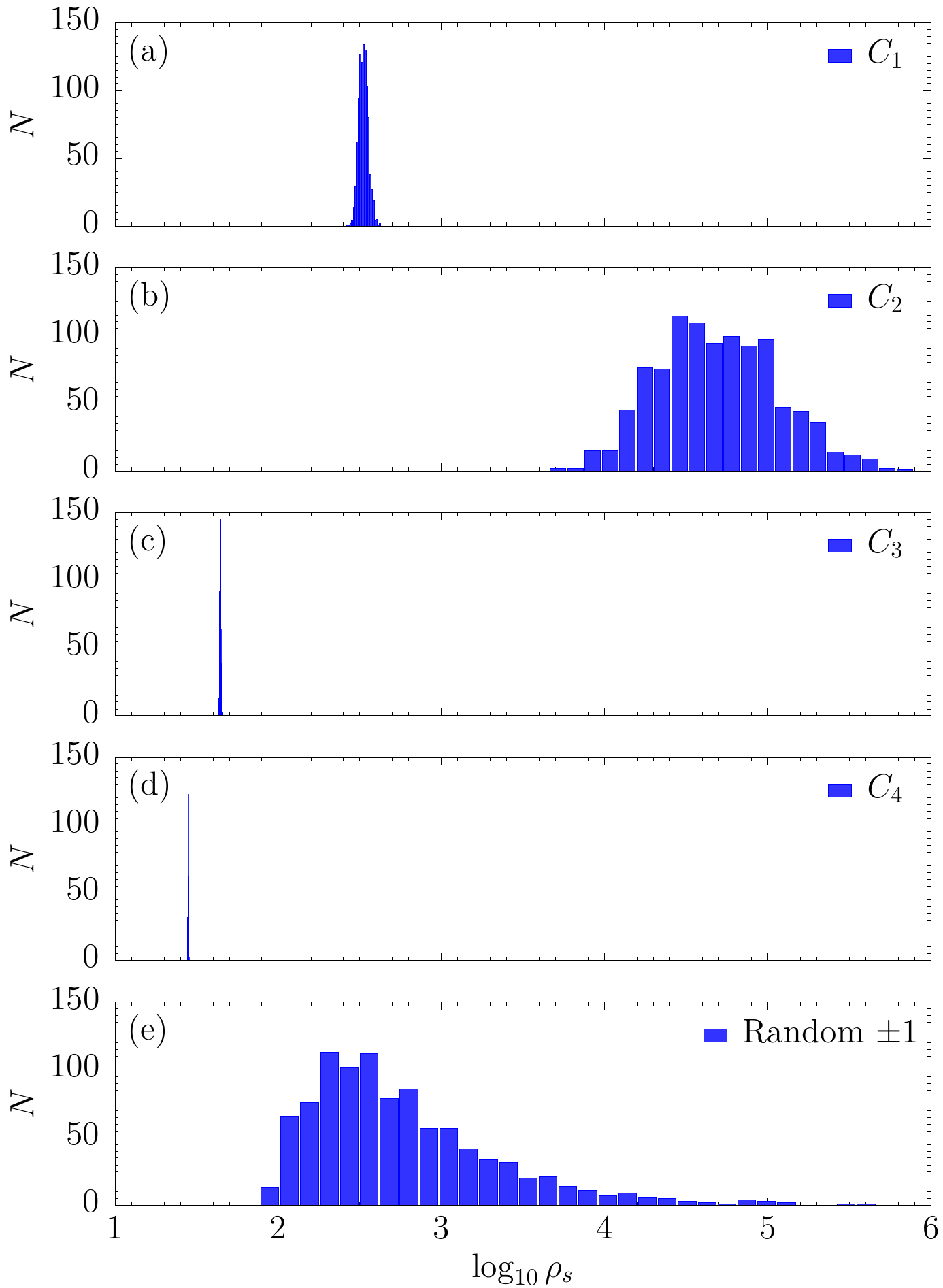} \caption{
  Distributions of the logarithm of the entropic family size,
    $\log_{10} \rho_s$, for planted problems with system size $L=32$
    constructed using subproblems drawn from a single subproblem type:
    (a) $C_1$, (b) $C_2$, (c) $C_3$, and (d) $C_4$. Panel (e) shows the
    distribution of $\log_{10} \rho_s$ for bimodal spin glasses with
    $J_{ij} \in \{\pm 1\}$. In each case, $1000$ instances per class
    were used.}
 \label{fig:hist_base_classes}
\end{figure}

We first focus on problems constructed using subproblems drawn from a
single subproblem class. Let us refer to the resulting four instance
classes as base classes. Figure \ref{fig:hist_base_classes} shows the
distributions of $\log_{10} \rho_s$ for the four base classes, along
with the results for random bimodal spin glasses, i.e., $J_{ij} \in
\{\pm 1\}$. For the bimodal instances, we use the same PAMC simulation
parameters as used for the base classes.  As
Fig.~\ref{fig:hist_base_classes} shows, $\rho_s$ has a very broad
distribution such that the distribution of $\log_{10} \rho_s$ is at
least similar to a normal shape, and we hence use $\log_{10} \rho_s$ as
our preferred form of the hardness metric.

Figure~\ref{fig:hist_base_classes} shows that both the $C_2$ base class
and the bimodal class have very broad distributions, with the bimodal
distribution being strongly skewed towards hard problem instances. In
contrast, the $C_3$ and $C_4$ base classes have quite narrow
distributions which appear to have vanishing overlap with the bimodal
case, clearly indicating that the problems belonging to these classes
are easier than the bimodal spin glasses. $C_4$ has the leftmost
distribution among all classes, and hence produces the easiest
instances, at least according to the $\rho_s$ metric.  On the other
hand, the distribution for $C_2$ lies the farthest to the right, only
overlapping with the tail of the bimodal distribution.  Thus, we
conclude that $C_2$ instances are at least as hard as the hardest
instances belonging to the bimodal class.  Finally, based on the
$\rho_s$ hardness metric, we can list the four base classes in sequence
according to increasing order of hardness as $C_4$, $C_3$, $C_1$, and
$C_2$. We will later demonstrate that by mixing the four subproblem
types one can construct problems with varying degrees of complexity,
with the level of hardness as measured by $\rho_s$ lying between those
of the $C_4$ and $C_2$ base classes.

\begin{figure}[tb!] 
\includegraphics[width=\linewidth, trim={0cm 0 0 0}, clip]{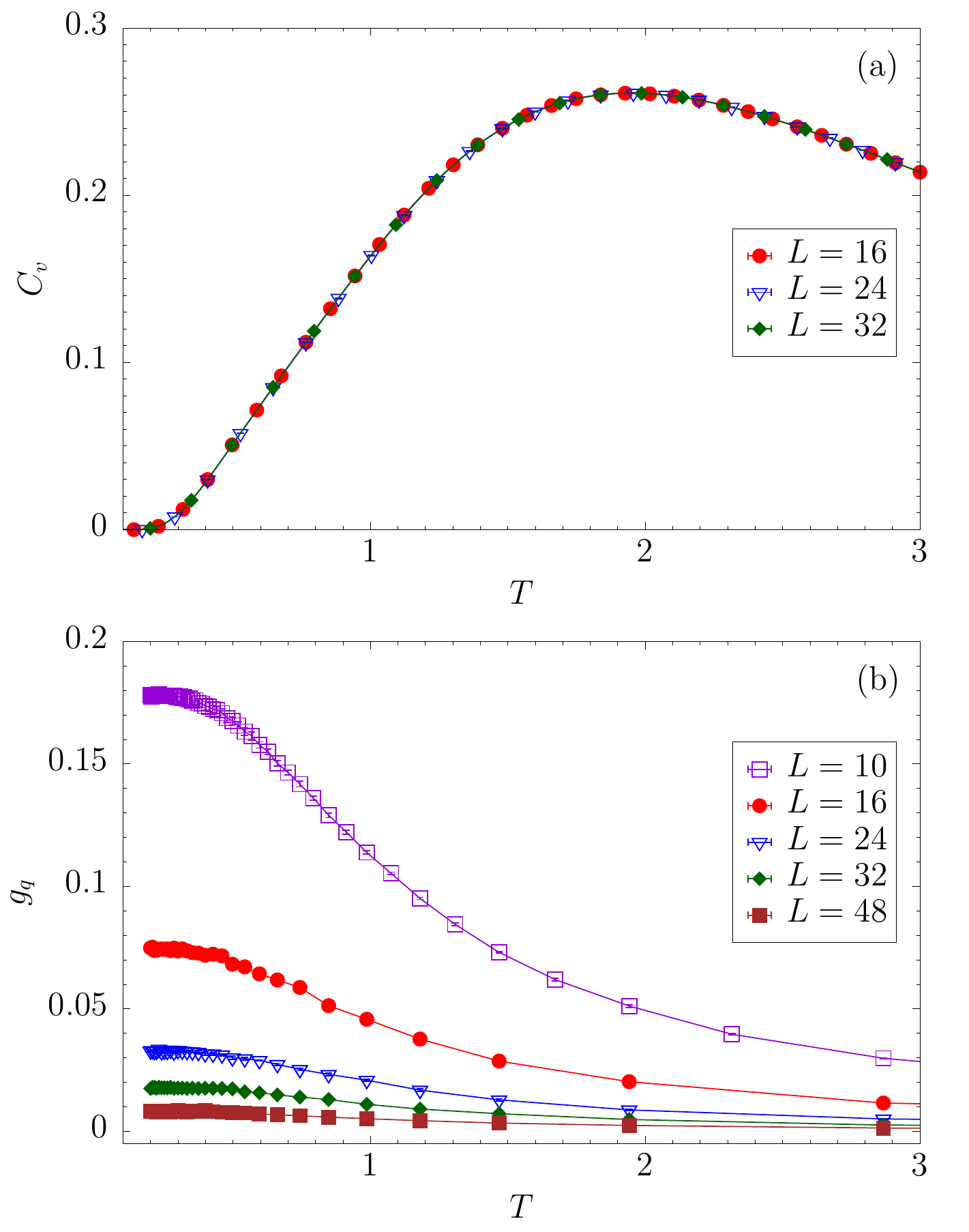}
\caption{ (a) The specific heat $C_v$ and (b) the Binder ratio $g_q$ as functions
    of temperature for the $C_3$ base class. The curves are for different system
    sizes. The size independence of the specific heat and lack of intersections of
    the Binder cumulants clearly indicate the absence of a thermodynamic phase
    transition.}
\label{fig:g_q_c3}
\end{figure}

\begin{figure}[tb!] 
\includegraphics[width=\linewidth, trim={0cm 0 0 0}, clip]{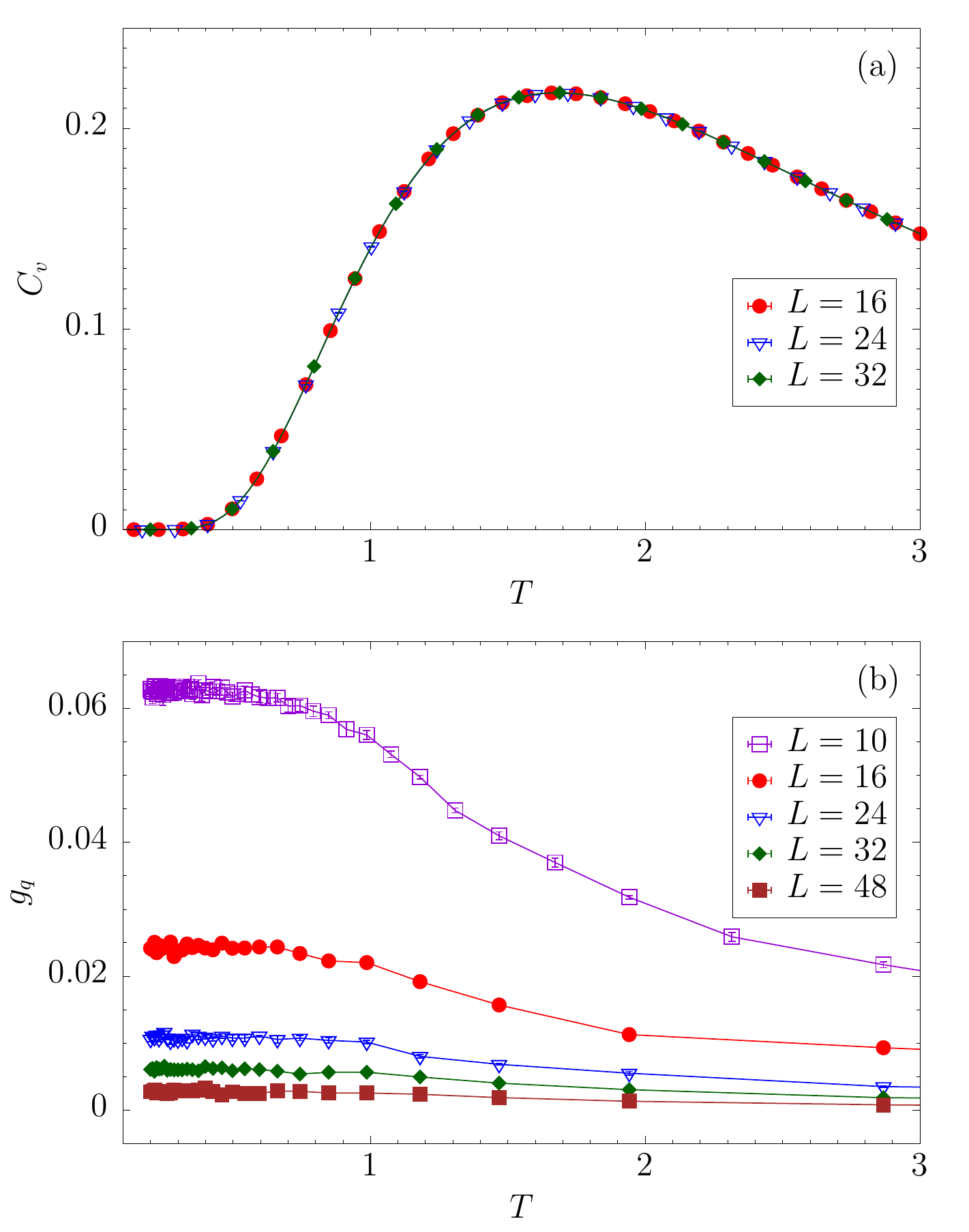}
\caption{
 (a) The specific heat $C_v$ and (b) the Binder ratio $g_q$ as functions
    of temperature for the $C_4$ base class. The curves are for different system
    sizes. The size independence of the specific heat and lack of intersections of
    the Binder cumulants clearly indicate the absence of a thermodynamic phase
    transition.
   }
\label{fig:g_q_c4}
\end{figure}

Before we proceed to problems constructed using mixtures of subproblem
types, let us investigate thermodynamic properties of the base classes
to uncover any noteworthy behavior such as the presence of phase
transitions. We first present results for the $C_3$ and $C_4$ classes
due to their trivial thermodynamic behavior in comparison to other base
classes.  Figure~\ref{fig:g_q_c3}(a) shows the specific heat $C_v$ and
Fig.~\ref{fig:g_q_c3}(b) shows the Binder ratio $g_q$ as a function of
temperature for the $C_3$ base class. In Fig.~\ref{fig:g_q_c4} we
present the same quantities for the $C_4$ base class. Here, the specific
heat is determined using the planar graph solver \cite{galluccio:00}
whereas the magnetization and the Binder ratio are measured using PAMC
simulations.  For the PAMC simulations, the results for system sizes
$L=16$, $24$, and $32$ were generated using the parameters presented in
Table~\ref{tab:sim_param}. For $L=10$, we use $R=1 \times 10^5$ and
$T=201$, and for $L=48$ we use $R=2 \times 10^5$ and $T=401$, with
$T_0=0.2$ and $N_S=10$ for both cases.  To ensure thermal equilibration,
we require the condition $R/\rho_s > 100$~\cite{wang:15e} to be met, and
when it is not satisfied by an instance it is rerun by increasing the
number of temperature steps $N_T$. As is demonstrated in
Figs.~\ref{fig:g_q_c3} and \ref{fig:g_q_c4}, the specific heat for both
$C_3$ and $C_4$ classes does not show any system-size dependence.
Moreover, the Binder cumulants $g_q$ for different system sizes do not
show any sign of an intersection, indicating the absence of
ferromagnetic or spin-glass transitions in these two base classes, which
hence show properties characteristic of disordered spin systems in
dimensions below the lower critical dimension. The implied absence of
critical slowing down and any significant barriers in the free-energy
landscape results in the relative ease of equilibrating the samples that
is reflected in the small values of $\rho_s$ shown in
Fig.~\ref{fig:hist_base_classes}.

\begin{figure}[tb!] 
\includegraphics[width=\linewidth, trim={0cm 0 0 0}, clip]{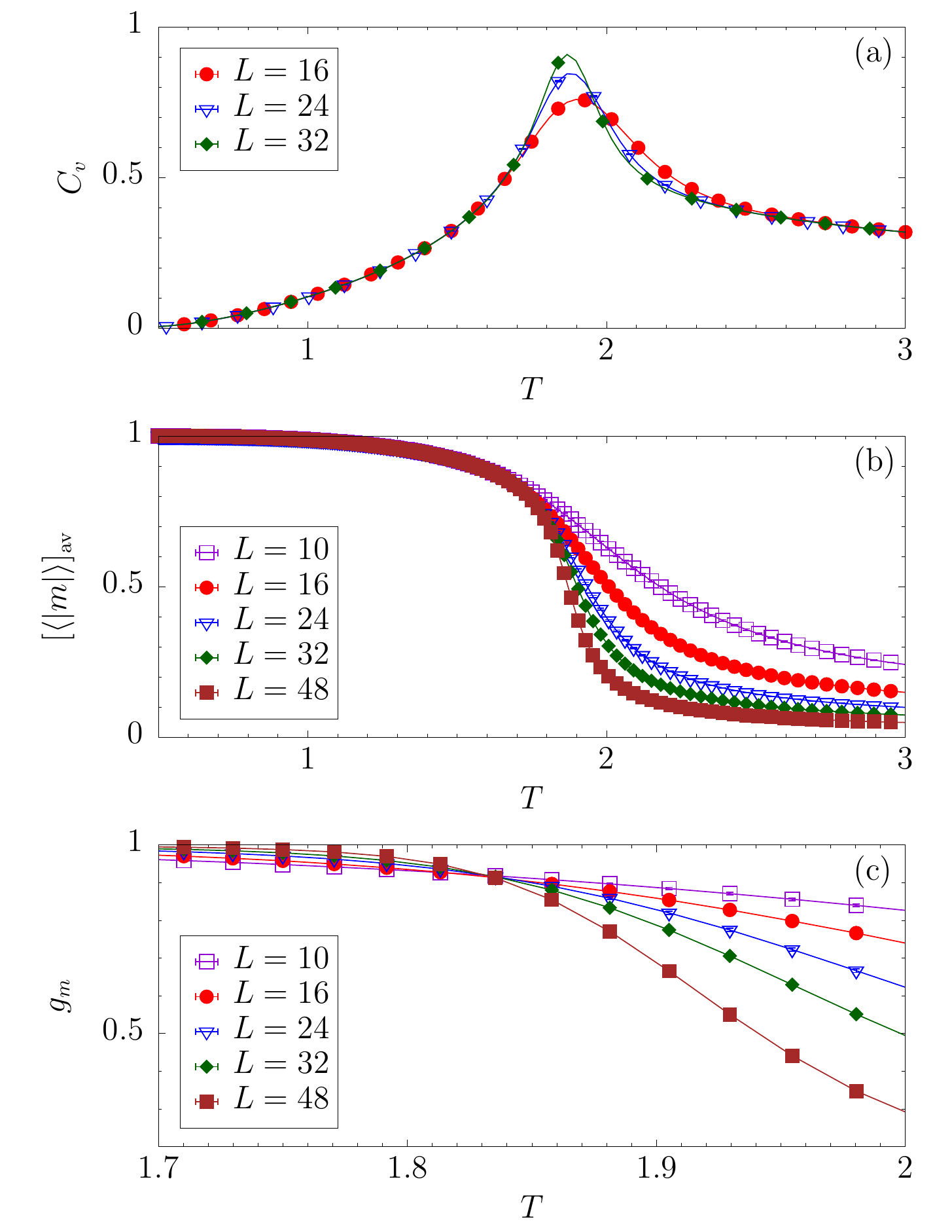}
\caption{ (a) Specific heat $C_v$, (b) average magnetization
    $[\langle |m| \rangle]_\text{av}$, and (c) Binder parameter $g_m$ as
    functions of temperature $T$ for the $C_1$ base class. The curves are for
    different system sizes. The curves for $g_m$ for different system sizes cross at
    a common finite temperature $T_c = 1.834(2)$, which indicates a ferromagnetic
    phase transition.}
\label{fig:g_m_c1}
\end{figure}

In Fig.~\ref{fig:g_m_c1}, we present results for the specific heat
$C_v$, the average magnetization $[\langle |m| \rangle]_\text{av}$, and
the Binder ratio $g_m$ for the $C_1$ base class.  The specific heat and
the average magnetization resemble that of the ferromagnetic Ising
model, with a logarithmic divergence of $C_v$ and the magnetization
scaling to zero at the transition point $\sim
(T-T_c)^{1/8}$~\cite{mccoy:73}. Moreover, the Binder parameter values $g_m$
for different system sizes clearly intersect at a common temperature
$T_c$. Using a scaling collapse~\cite{katzgraber:06}, we determine the
critical temperature to be $T_c = 1.834(2)$, with $\nu_m = 1.001(5)$.
Our results indicate that despite being a disordered and frustrated
system the $C_1$ base class has predominantly ferromagnetic features.
This ferromagnetic ordering is a necessary consequence of the fact that
these instances only have a single ground state (excluding spin-flip
symmetry), which is the planted ferromagnetic solution.

\begin{figure}[tb!] 
\includegraphics[width=\linewidth, trim={0cm 0 0 0}, clip]{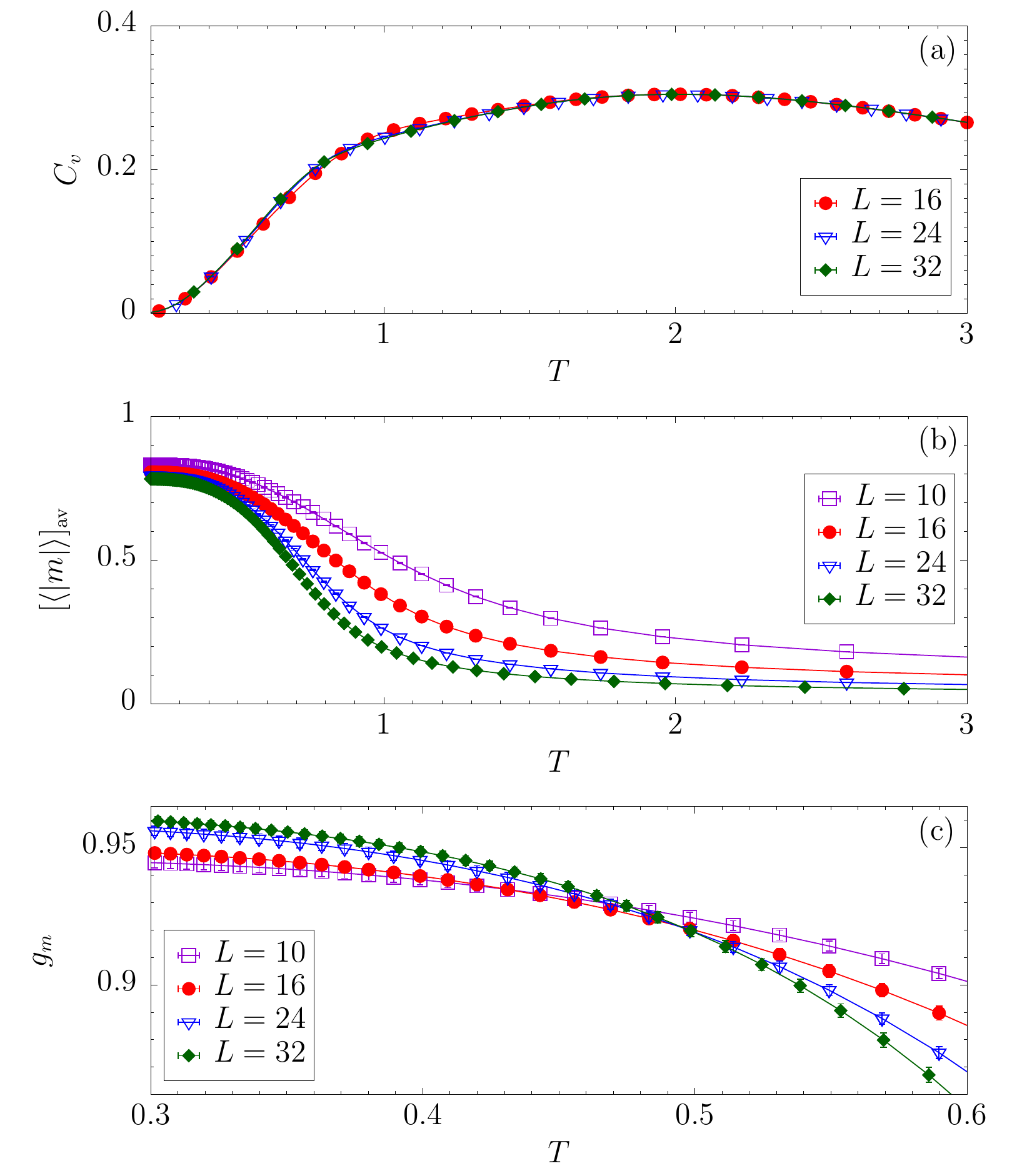}
\caption{ (a) Specific heat $C_v$, (b) average magnetization
    $[\langle |m| \rangle]_\text{av}$, and (c) Binder parameter $g_m$ as
    functions of temperature $T$ for the $C_2$ base class.  The curves are for
    different system sizes.
    $g_m$ for different system sizes intersect in the temperature interval 
    $T \in \left(0.4, 0.55\right)$, which indicates ferromagnetic ordering.
    }
\label{fig:g_m_c2}
\end{figure}

Finally, we explore properties of the $C_2$ base class.  In
Fig.~\ref{fig:g_m_c2}, we show the specific heat $C_v$, magnetization
$[\langle |m| \rangle]_\text{av}$, and Binder parameter $g_m$ for the
$C_2$ class. 
Here, system sizes beyond $L=32$ were not simulated due to difficulties
in thermalizing $C_2$ instances. 
Figure \ref{fig:g_m_c2}(c) shows that $g_m$ for different
system sizes intersect in the temperature interval $T \in \left(0.4,
0.55\right)$, which indicates that some ferromagnetic ordering occurs.
The fact that the curves do not intersect at a common point implies
that corrections to scaling are significant for these system sizes.
Hence, we do not attempt to extract the critical temperature via a
scaling analysis. The average magnetization [Fig.~\ref{fig:g_m_c2}(b)]
also shows features reminiscent of a ferromagnetic transition, although,
as $T \rightarrow 0$, $[\langle |m| \rangle]_\text{av}$ appears to
approach a value close to $0.8$, rather than $1.0$ as in the case of the
ferromagnetic Ising model.  The specific heat [Fig.~\ref{fig:g_m_c2}(a)]
does not show a discernible system-size dependence, at least for the
system sizes considered.

\begin{figure}[tb!] 
\includegraphics[width=0.8\linewidth, trim={0cm 0 0 0}, clip]{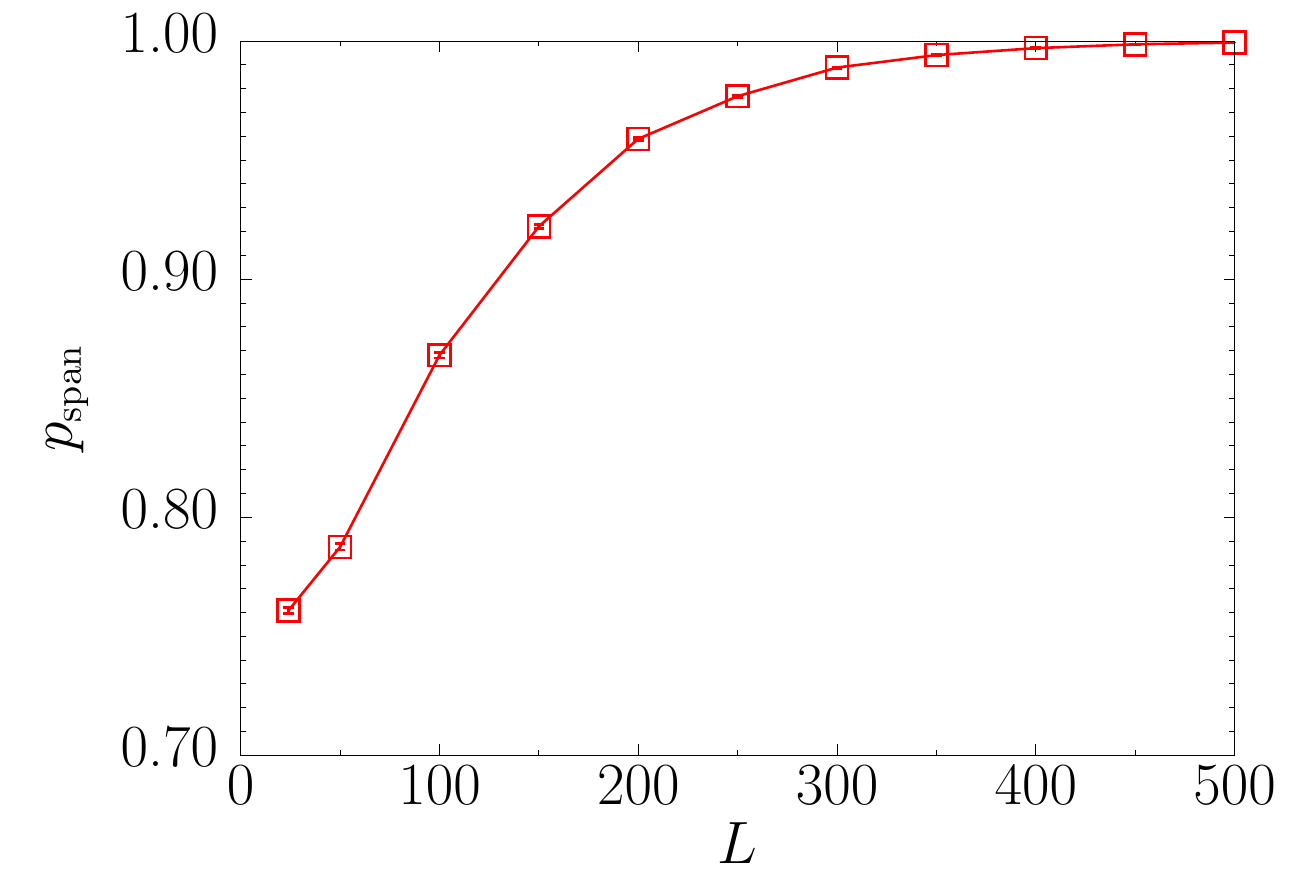}
\caption{Spanning probability $p_\text{span}$ of strong-ferromagnetic (+2)
		bond clusters in $C_2$ base class instances, plotted against system size $L$.
		For each system size, $p_\text{span}$ was estimated using 100,000 instances.
		For simplicity, free boundary conditions were used.
		The results show that $p_\text{span} \rightarrow 1$ with increasing $L$.
}
\label{fig:p_span}
\end{figure}

\begin{figure}[tb!] 
\includegraphics[width=0.5\linewidth, trim={0cm 0 0 0}, clip]{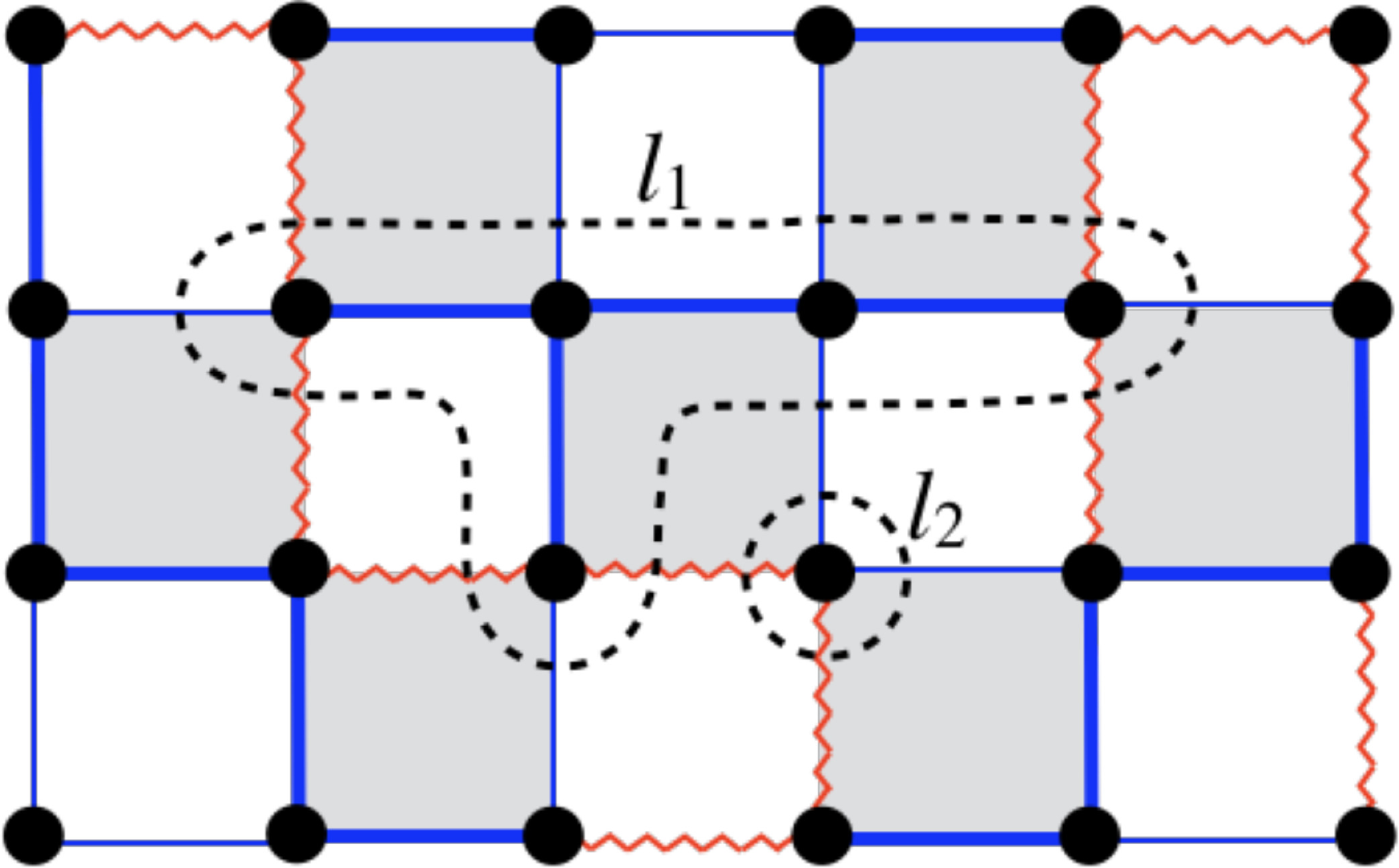}
\caption{A section of the lattice structure of a $C_2$ base class instance.  Red
    wiggly lines represent antiferromagnetic couplers with value $-1$, whereas
    straight blue lines represent ferromagnetic couplers with values $+1$ (thin
    lines) and $+2$ (thick lines), respectively. The loop $l_1$ encloses a cluster of
    strong-ferromagnetic ($+2$) bonds, while $l_2$ encloses an ``isolated'' spin with
    no $+2$ bonds attached.  Due to intrinsic properties of the $C_2$ lattice
    structure, $+2$-bond clusters and isolated spins in $C_2$ instances have equal
    number of $+1$ and $-1$ bonds attached.  Thus, the flipping of an isolated spin,
    or the simultaneous flipping of all the spins in a $+2$-bond cluster, preserves
    energy.}
\label{fig:bond_clusters}
\end{figure}

\begin{figure}[tb!] 
\includegraphics[width=\linewidth, trim={0cm 0 0 0}, clip]{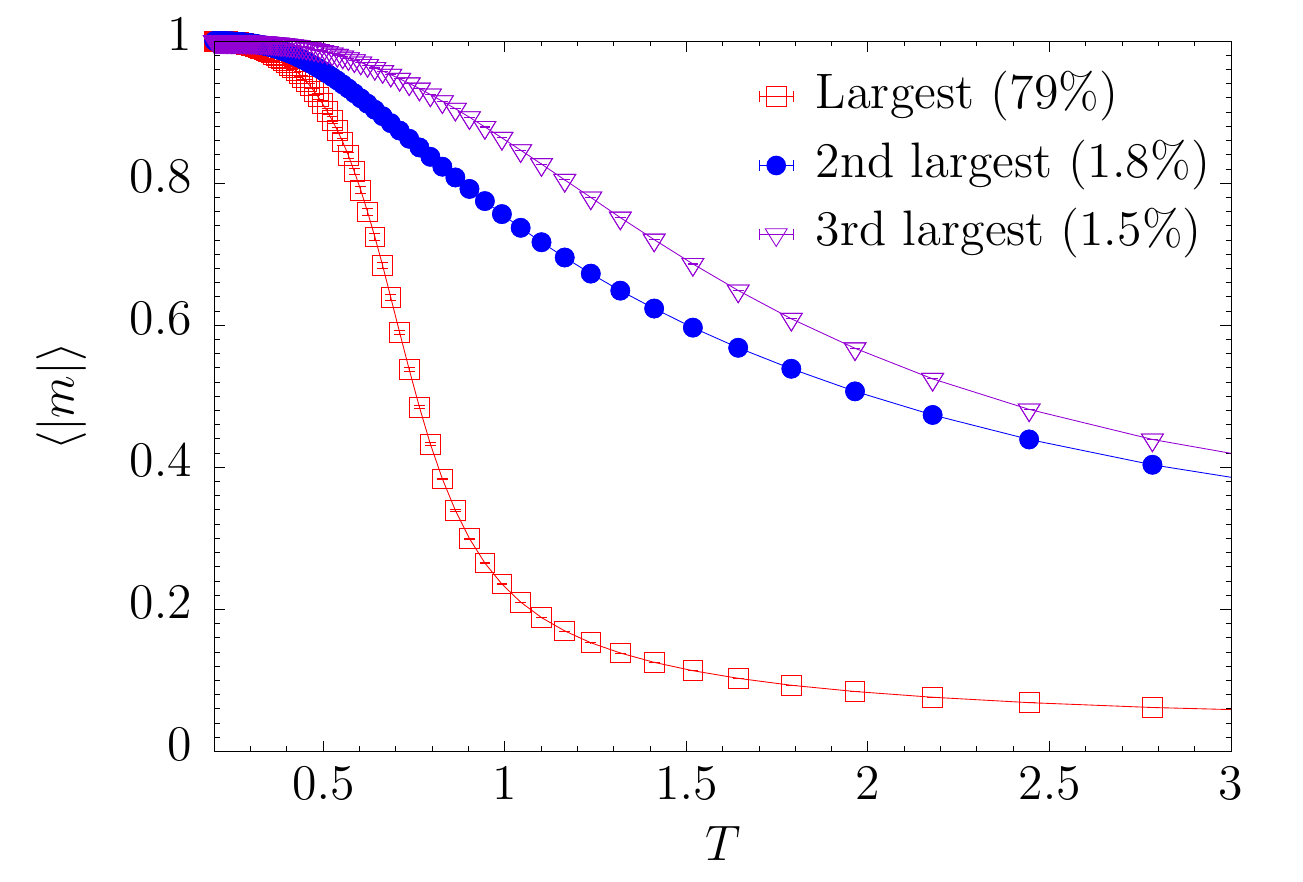}
\caption{ Average magnetization $\langle |m| \rangle$ as a function of temperature
    for the three largest clusters of strong-ferromagnetic ($+2$) bonds of a single
    $C_2$ base class instance ($L=32$). The largest cluster contains 79\% of the
    spins in the lattice, while the second and third largest clusters, respectively,
    contain 1.8 and 1.5\% of the spins.  }
\label{fig:m_clusters}
\end{figure}

This peculiar ordering behavior can be understood by considering the
clustering behavior of strong-ferromagnetic ($+2$) bonds in $C_2$
instances. In $C_2$ instances, $50\%$ of the couplers are $+2$ bonds.
By investigating the percolation properties of $+2$ bonds, we find that
the spanning probability asymptotically approaches unity in the
thermodynamic limit (see Fig.~\ref{fig:p_span}). This indicates that the lattice is in the
percolating phase, and that in the thermodynamic limit there is a
nonvanishing fraction of sites in the largest cluster of $+2$ bonds.  We
find that for the lattice sizes considered approximately 80\% of the
spins belong to the largest bond cluster. Note that this is in contrast
to an {\em uncorrelated\/} percolation problem in its ordered phase,
where close to 100\% of lattice sites would belong to the percolating
cluster.  The remaining spins either belong to smaller $+2$-bond
clusters scattered around the lattice or they are isolated sites with
no $+2$ bonds attached (see Fig.~\ref{fig:bond_clusters} for an
illustration).  Due to intrinsic properties of the $C_2$ lattices,
$+2$-bond clusters and ``isolated'' spins have equal number of
weak-ferromagnetic ($+1$) and weak-antiferromagnetic ($-1$) bonds
attached.  As a result, the flipping of an isolated spin, or the
simultaneous flipping of all the spins in a $+2$-bond cluster, leaves
the energy unchanged.  By examining the ground-state spin
configurations, we find that all $+2$ bonds are satisfied in the ground
state, that is, spins within each cluster are aligned with one another.
Thus, the ground-state degeneracy in $C_2$ instances arises from the
energy-free flipping of entire clusters or isolated spins. As the
critical temperature is approached from above, each cluster would
undergo a transition from a randomly oriented (paramagnetic) state to a
ferromagnetically aligned state.  This is demonstrated in
Fig.~\ref{fig:m_clusters}, where we show the average magnetization
$\langle |m| \rangle$ as a function of temperature for the three largest
clusters of a single $C_2$ instance. As $T \rightarrow 0$, we observe
that $\langle |m| \rangle \rightarrow 1$ as the spins within each
cluster are ferromagnetically aligned.

A main factor leading to the observed hardness in $C_2$ instances is the
critical slowing down that occurs as the system undergoes a
ferromagnetic transition at a considerably low temperature $T \in
\left(0.4, 0.55\right)$.  We point out that cluster
algorithms~\cite{swendsen:87,wolff:89} could be effective in reducing
critical slowing down and improving the performance.  However, an
analysis on the performance of problem-specific algorithms is beyond the
scope of this paper, and hence was not attempted. The
ferromagnetic nature of $C_1$ and $C_2$ base-class instances is not
surprising when considering the fraction of strong-ferromagnetic bonds
in $C_1$ and $C_2$ plaquettes.  Three out of four bonds in a $C_1$
plaquette are strong-ferromagnetic bonds, and, therefore, $C_1$
instances have the strongest ferromagnetic properties among all classes.
With the replacement of each strong-ferromagnetic bond with a
weak-ferromagnetic bond, the plaquettes progressively become less
ferromagnetic and more frustrated in the order $C_1 \rightarrow C_2
\rightarrow C_3 \rightarrow C_4$. While both $C_1$ and $C_2$ order
ferromagnetically with the accompanying critical slowing down, $C_2$
also shows a significant ground-state degeneracy
[see Eq.~(\ref{eq:degeneracies})], and the expected energy barriers
between the multitude of ground states lead to additional slow dynamics
at low temperatures that is absent for $C_1$, explaining the much weaker
computational challenge provided by $C_1$ instances as compared to $C_2$
samples.

\subsection{Mixtures of two subproblem types} \label{sec:two_classes}

\begin{figure}[tb!]
  \includegraphics[width=\linewidth, trim={1.5cm 0 0 0}, clip]{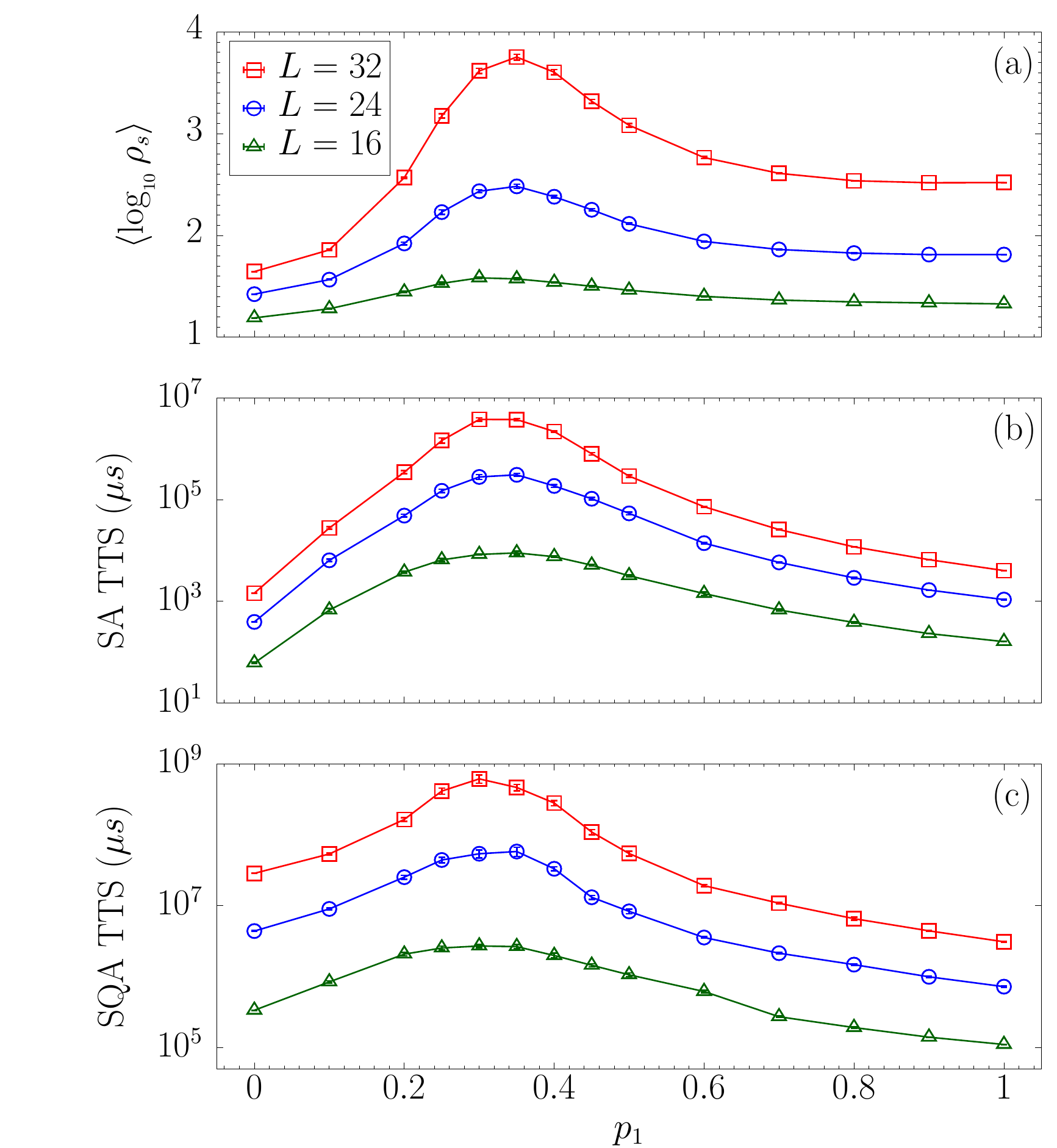}
  \caption{ (a) Population annealing $\langle \log_{10} \rho_s \rangle$, (b)
    simulated annealing optimal median TTS, and (c) simulated quantum annealing
    optimal median TTS for instance classes composed of mixtures of $C_1$ and $C_3$
    subproblem types. The results are plotted against $p_1$, which is the
    probability of choosing subproblems from class $C_1$.  The curves are for three
    system sizes, $L=16$, $24$, and $32$.  $\langle \log_{10} \rho_s \rangle$ and
    median TTS values are estimated using $200$ problems per instance class, per
    system size. All panels have the same horizontal scale.}
 \label{fig:peak_c1_c3}
\end{figure}

\begin{figure}[tb!] 
  \includegraphics[width=\linewidth, trim={1.5cm 0 0 0}, clip]{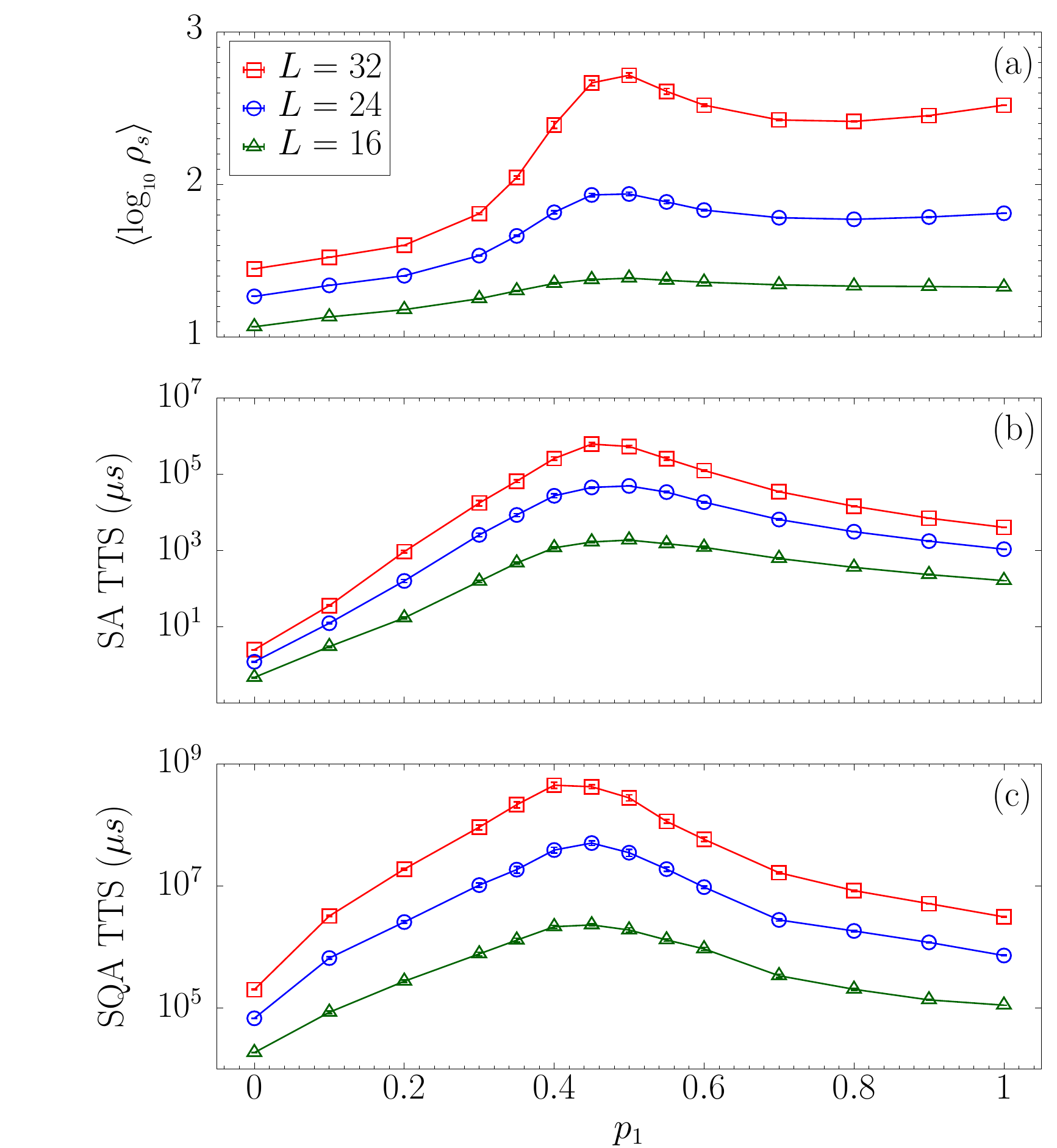}
  \caption{ (a) Population annealing $\langle \log_{10} \rho_s \rangle$, (b)
    simulated annealing optimal median TTS, and (c) simulated quantum annealing
    optimal median TTS for instance classes composed of mixtures of $C_1$ and $C_4$
    subproblem types.  The results are plotted against $p_1$, which is the
    probability of choosing subproblems from class $C_1$.  The curves are for three
    system sizes, $L=16$, $24$, and $32$.  $\langle \log_{10} \rho_s \rangle$ and
    median TTS values are estimated using $200$ problems per instance class, per
    system size. All panels have the same horizontal scale.}
 \label{fig:peak_c1_c4}
\end{figure}

We now investigate planted instances constructed using subproblems drawn
from two plaquette types.  We first focus on two regimes where
interesting variations in hardness can be observed: mixtures of $C_1$
and $C_3$ plaquettes, and mixtures of $C_1$ and $C_4$ plaquettes.
Rather than comparing $\log_{10} \rho_s$ distributions, here we consider
the disorder averages $\langle \log_{10} \rho_s \rangle$ computed over
$200$ problems drawn from each instance class per system size.
Figure~\ref{fig:peak_c1_c3}(a) shows $\langle \log_{10} \rho_s \rangle$
for $C_1$--$C_3$ plaquette mixtures plotted as a function of $p_1$,
which is the probability of choosing subproblems from class $C_1$. Thus,
$p_1=1$ corresponds to the instance class with all subproblems drawn
from $C_1$ plaquette type ($C_1$ base class), whereas $p_1=0$ represents
instances with all $C_3$ subproblems ($C_3$ base class). Results are
shown for three system sizes, $L=16$, $24$, and $32$. Figures
~\ref{fig:peak_c1_c3}(b) and ~\ref{fig:peak_c1_c3}(c) show the optimized
median TTS for SA and SQA, again estimated using $200$ problem instances.  Figure
\ref{fig:peak_c1_c4} shows the same quantities for $C_1$--$C_4$
plaquette mixtures. Results from all three algorithms clearly indicate
peaks in problem hardness at $p_1 \approx 0.35$ for $C_1$--$C_3$
mixtures, and $p_1 \approx 0.45$ for $C_1$--$C_4$ mixtures, i.e., an
easy-hard-easy transition. This easy-hard-easy transition is akin to
the phase transition one observes in Boolean satisfiability ($k$-SAT)
problems at a certain clauses-to-variables ratio, below which the
formula is satisfiable and above which it is
unsatisfiable~\cite{bollobas:01a}.  
However, we point out that our problems are exactly solvable in polynomial
time, and therefore the observed hardness transition is more analogous to
the transition in 2-SAT problems than the $k>0$ case for
which one observes a discontinuous transition.
A similar hardness peak has been
observed for the frustrated loop problems on the Chimera lattice as the
loop density is varied~\cite{hen:15a}.

\begin{figure}[h] 
\includegraphics[width=\linewidth, trim={0cm 0 0 0}, clip]{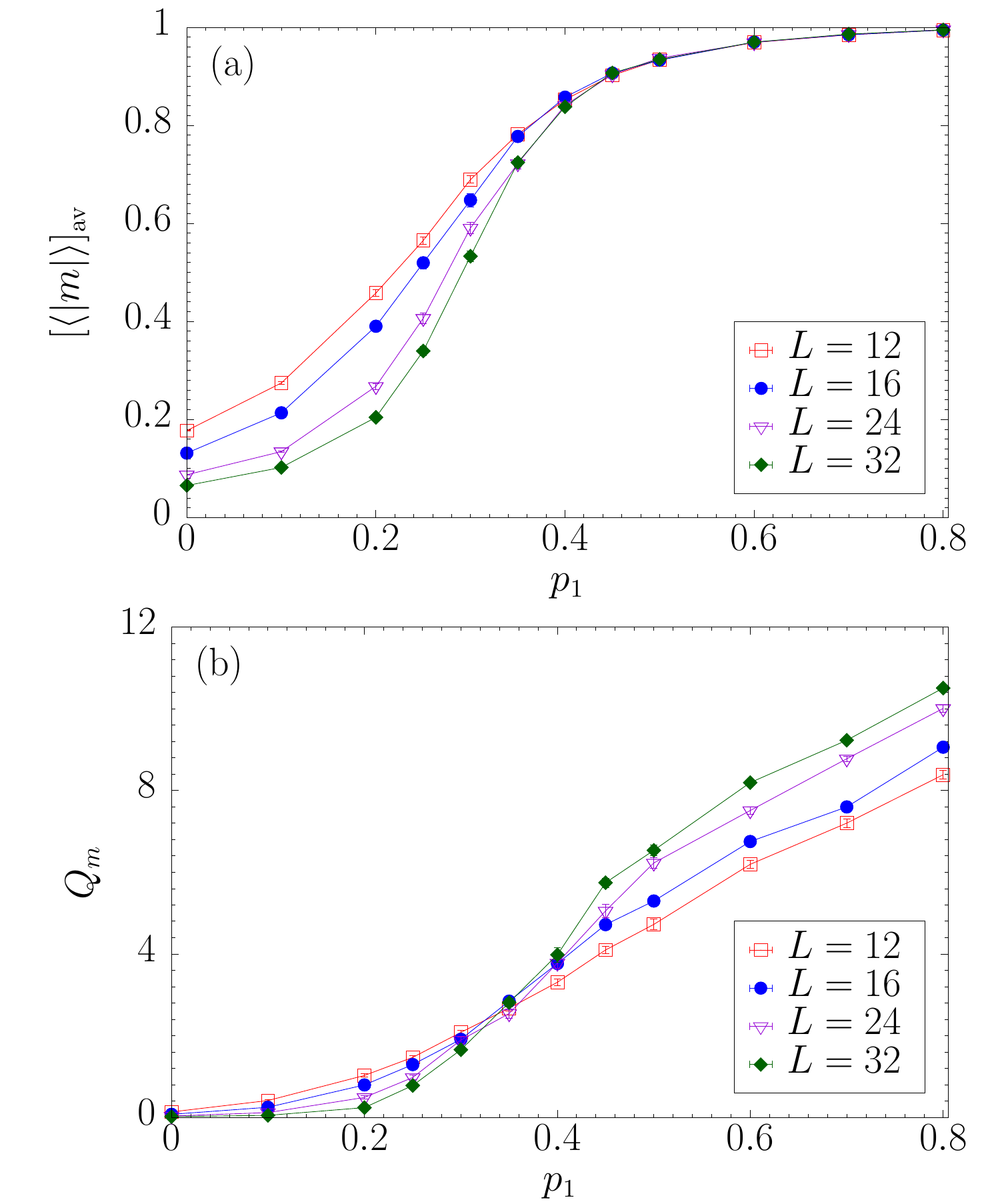}
\caption{
(a) Average magnetization $[\langle |m| \rangle]_\text{av}$ and (b) $Q_m
= -\ln(1-g_m)$ for $C_1$--$C_3$ instance classes, measured at the lowest
temperature simulated, $T_\text{min} = 0.2$. The horizontal axis
represents $p_1$.  }
\label{fig:ferro_transition_c1_c3}
\end{figure}

It is interesting to explore whether these hardness peaks coincide with
or are driven by thermodynamic phase transitions.  Based on the
thermodynamic behavior of the instance classes at $p_1=0$ ($C_3$ and
$C_4$) and $p_1=1$ ($C_1$), it is reasonable to speculate that at zero
temperature, as $p_1$ is varied from $0$ to $1$, the system may undergo
a transition from a disordered phase to a ferromagnetic phase.  To
further explore this hypothesis, we focus on the $C_1$--$C_3$ plaquette
mixtures and measure $[\langle |m| \rangle]_\text{av}$ and $g_m$ at the
lowest temperature $T_\text{min} = 0.2$ for different values of $p_1 \in
[0,1]$. The results are shown in Fig.~\ref{fig:ferro_transition_c1_c3} as a
function of $p_1$, where Fig.~\ref{fig:ferro_transition_c1_c3}(a) shows
$[\langle |m| \rangle]_\text{av}$ and Fig.~\ref{fig:ferro_transition_c1_c3}(b)
shows the quantity $Q_m = -\ln(1-g_m)$. 
System sizes beyond $L=32$ were not simulated due to difficulties
in thermalizing instance classes in the vicinity of the hardness transition. 
Here, the quantity $Q_m =
-\ln(1-g_m)$ was chosen over $g_m$ as it has a smaller curvature near
the transition~\cite{moebius:09b}, allowing for a more accurate
estimation of the crossing point.  Figure \ref{fig:ferro_transition_c1_c3}(a)
shows that $[\langle |m| \rangle]_\text{av} \rightarrow 1$ as $p_1
\rightarrow 1$, consistent with the ferromagnetic properties of the
$C_1$ base class.  As Fig.~\ref{fig:ferro_transition_c1_c3}(b) shows, $Q_m$
for different system sizes intersect at a nonzero $p_1$ value, further 
confirming the presence of a ferromagnetic transition.  
Assuming a scaling behavior of the form $Q_m \sim F [L^{1/\kappa} (p_1-p_1^c)]$, we
perform a finite-size scaling analysis to determine the point of intersection
$p_1^c$.  From finite-size scaling we expect $\kappa = \nu$, and universality implies
$\nu = 1$ as we are dealing with an Ising transition in two dimensions. This leads to
an acceptable fit and the estimate $p_1^c=0.352(7)$.

\begin{figure}[h] 
\includegraphics[width=\linewidth, trim={0cm 0 0 0}, clip]{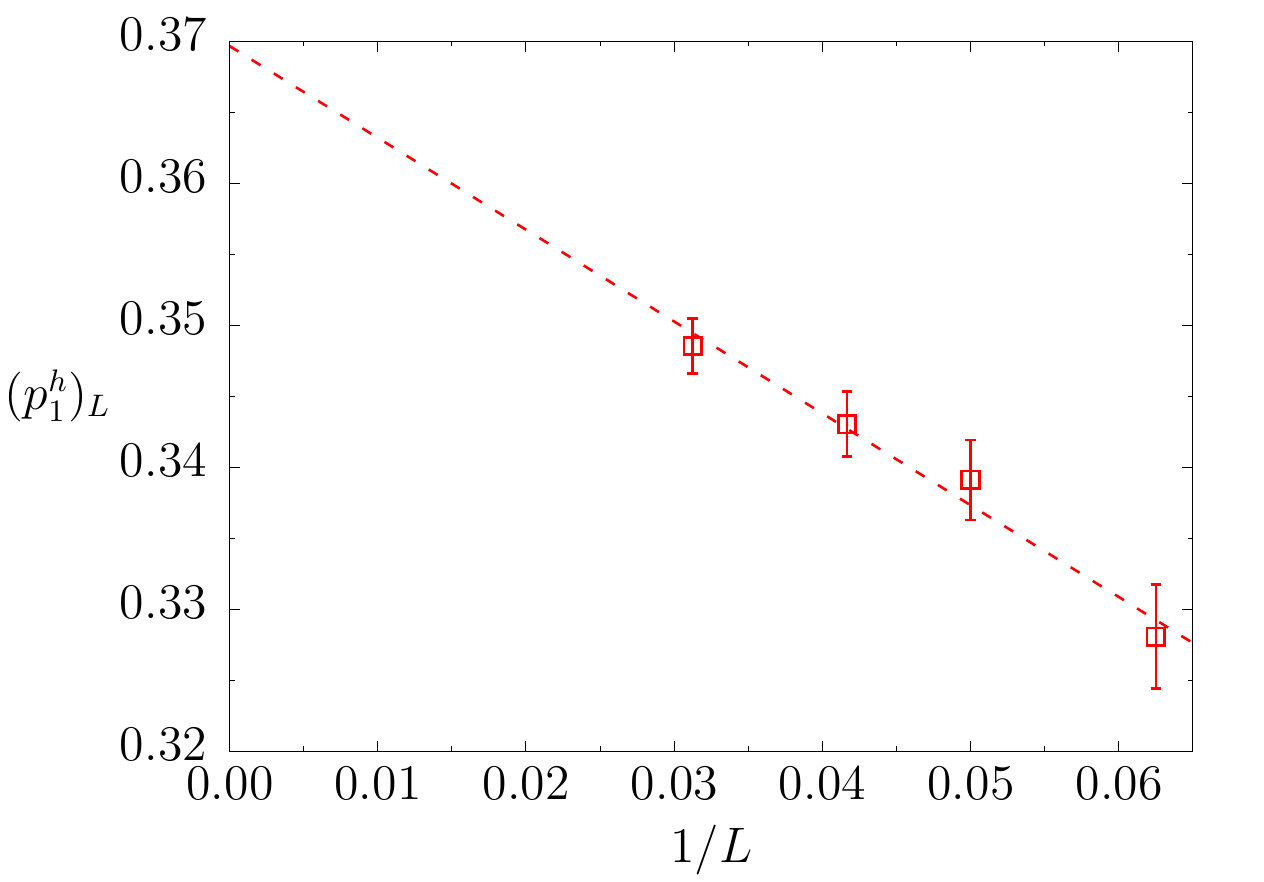}
\caption{
The position of the hardness peak in $C_1-C_3$ mixtures obtained using 
$\langle \log_{10} \rho_s \rangle$ data
for different system sizes, plotted against $1/L$.
The data points are for the system sizes $L=16$, $20$, $24$, and $32$.
The dashed line is a fit to $(p_1^h)_L = A + \lambda/L$,
with $A=0.370(6)$ and $\lambda=-0.6(1)$.
The fit provides a rough estimate of the asymptotic value of the peak position
in the limit $L \rightarrow \infty$, $p_1^h = A = 0.370(6)$.
}
\label{fig:asymptotic_peak}
\end{figure}

A visual inspection of the hardness curves 
in Fig.~\ref{fig:peak_c1_c3} shows that the peak in hardness 
occurs in the vicinity of this transition point. To obtain a rough estimate 
of the asymptotic value of the hardness peak position as $L \rightarrow \infty$, 
for each $L$, we fit cubic polynomials to the data points of 
$\langle \log_{10} \rho_s \rangle$ close to the peak, and estimate the 
peak positions $(p_1^h)_L$. Fig.~\ref{fig:asymptotic_peak} shows $(p_1^h)_L$ 
plotted against $1/L$. By fitting a function of the form $(p_1^h)_L = A + \lambda/L$, 
we determine the asymptotic peak position to be $p_1^h = A = 0.370(6)$, 
where we have estimated the error bar using bootstrap resampling. 
The close proximity of the transition point $p_1^c$ to the asymptotic hardness
peak position $p_1^h$ suggests that the hardness transition
is, indeed, driven by the ferromagnetic transition.

\begin{figure}[h] 
\includegraphics[width=\linewidth, trim={0cm 0 0 0}, clip]{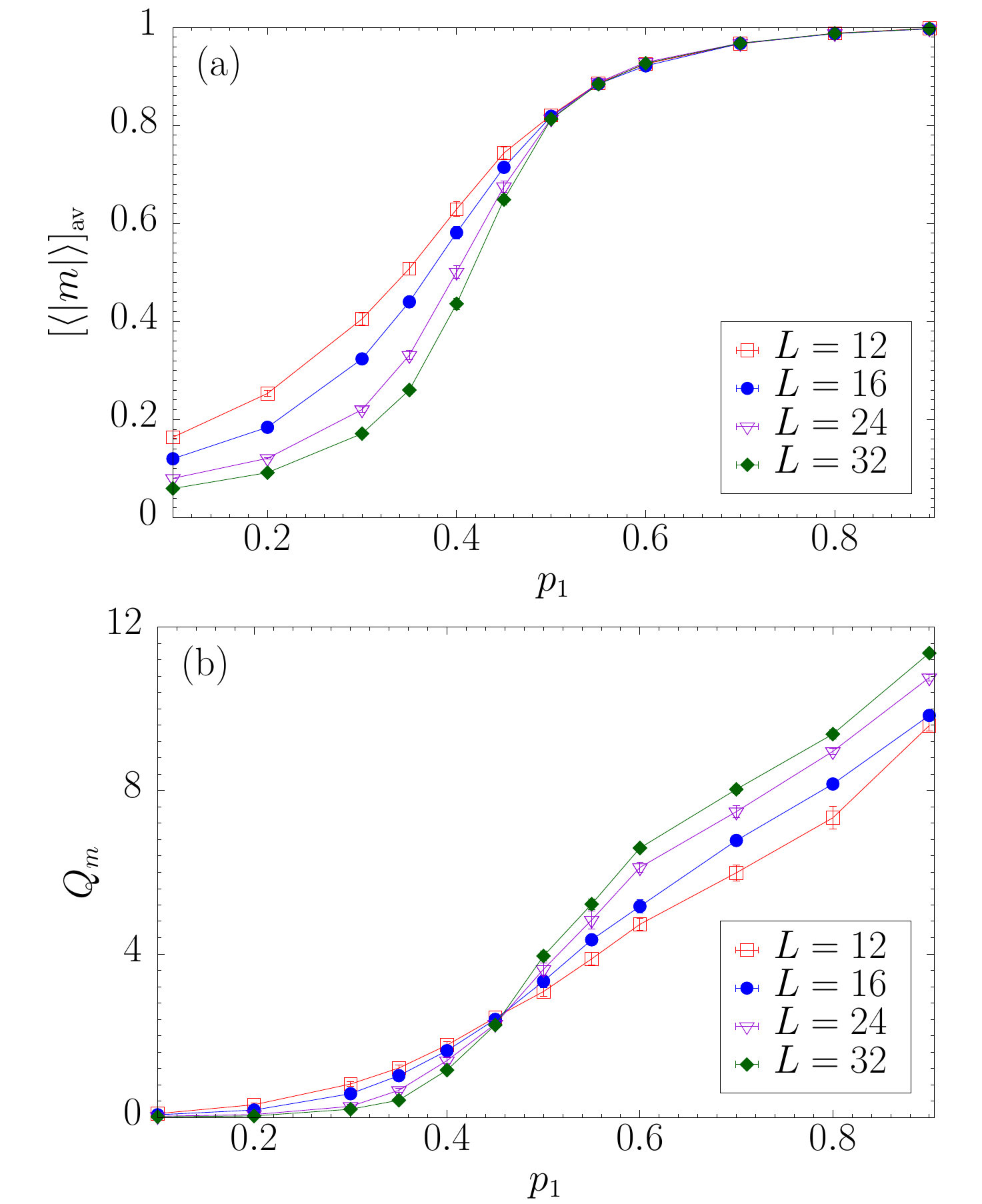}
\caption{
(a) Average magnetization $[\langle |m| \rangle]_\text{av}$ and (b) $Q_m
= -\ln(1-g_m)$ for $C_1$--$C_4$ instance classes, measured at the lowest
temperature simulated, $T_\text{min} = 0.2$. The horizontal axis
represents $p_1$.  }
\label{fig:ferro_transition_c1_c4}
\end{figure}

We now turn to the thermodynamic properties of $C_1$--$C_4$
mixtures. Fig.~\ref{fig:ferro_transition_c1_c4}(a) and 
Fig.~\ref{fig:ferro_transition_c1_c4}(b), respectively, show 
$[\langle |m| \rangle]_\text{av}$ and $Q_m$ as functions of $p_1$ for 
$C_1$--$C_4$ mixtures. Both $[\langle |m| \rangle]_\text{av}$ and $Q_m$ behave 
in qualitatively similar ways to the corresponding quantities for $C_1$--$C_3$ mixtures. 
From a finite-size scaling analysis, we estimate the intersection point in $Q_m$ 
to be $p_1^c = 0.464(6)$.
An inspection of the hardness curves in Fig.~\ref{fig:peak_c1_c4}
indicates that the peak in hardness occurs in the proximity of this transition point. 
By investigating the scaling behavior of the peak position as a function of $1/L$,
we obtain a rough estimate of its asymptotic value, $p_1^h = 0.50(1)$.  
The consistent closeness between the intersection points in $Q_m$
and the hardness peak positions
corroborates that in both $C_1$--$C_3$ and $C_1$--$C_4$ mixtures 
the hardness transitions are driven by magnetic ordering.

It is noteworthy to mention that apart from the thermodynamic
properties we have also analyzed some heuristic rules based on local
measures of frustration~\cite{kobe:95}, but these quantities proved not
to be predictive of the hardness transitions or the relative levels of
hardness of the base classes~\cite{miyazaki:13}.

\begin{figure}[h] 
  \includegraphics[width=0.95\linewidth, trim={1.0cm 0 0 0}, clip]{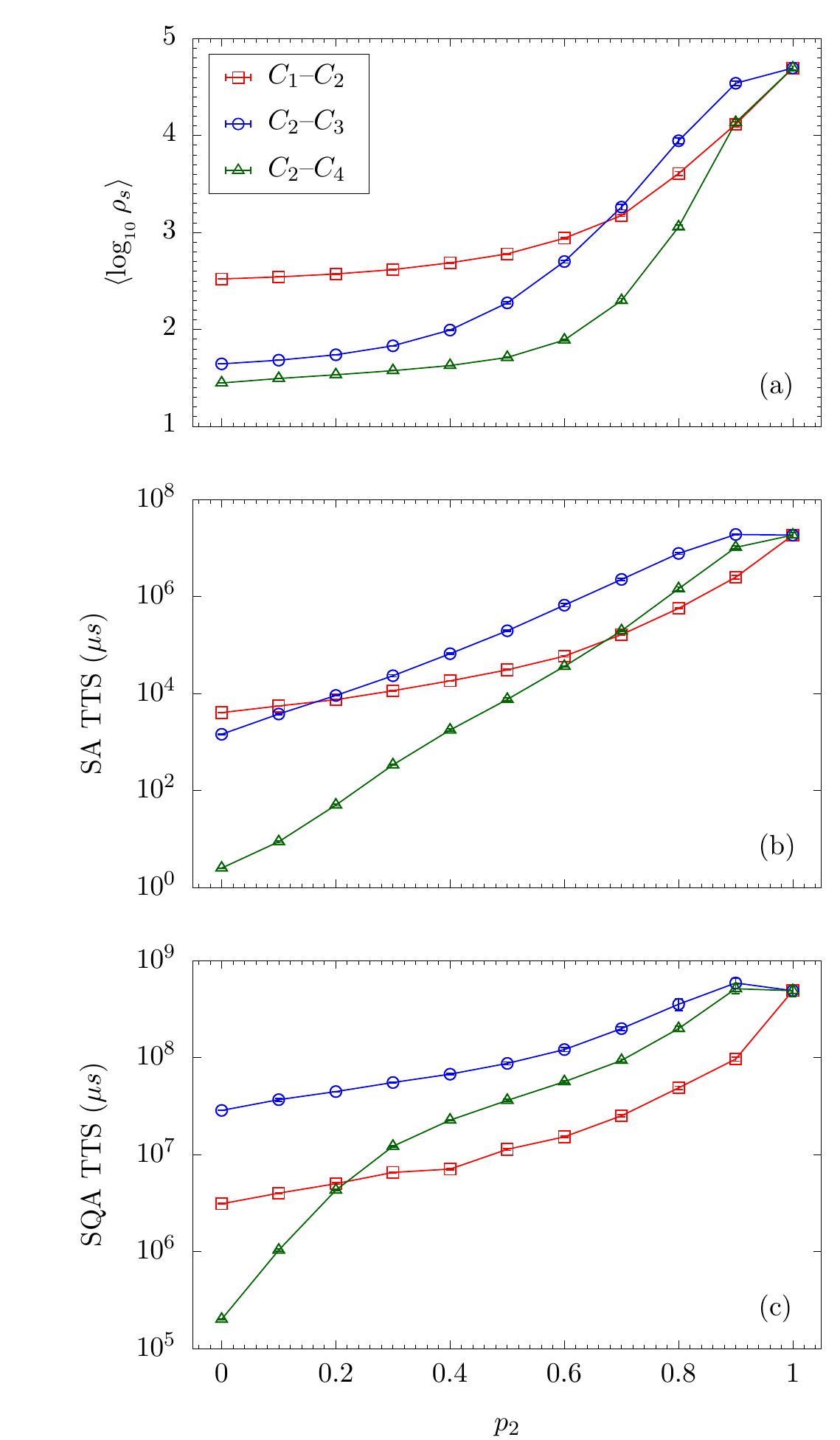}
  \caption{ (a) Population annealing $\langle \log_{10} \rho_s \rangle$, (b)
    simulated annealing optimal median TTS, and (c) simulated quantum annealing
    optimal median TTS for instance classes constructed by mixing $C_2$ subproblems with 
    one of the other subproblem types. 
    The results are for the system size $L=32$.
    The results are plotted against $p_2$, which is the
    probability of choosing subproblems from class $C_2$. $\langle \log_{10} \rho_s \rangle$ and
    median TTS values are estimated using $200$ problems per instance class, per
    system size. All panels have the same horizontal scale.}
 \label{fig:c2_binary_mixtures}
\end{figure}

\begin{figure}[h] 
\hspace*{-3.0em}\includegraphics[width=1.2\columnwidth]{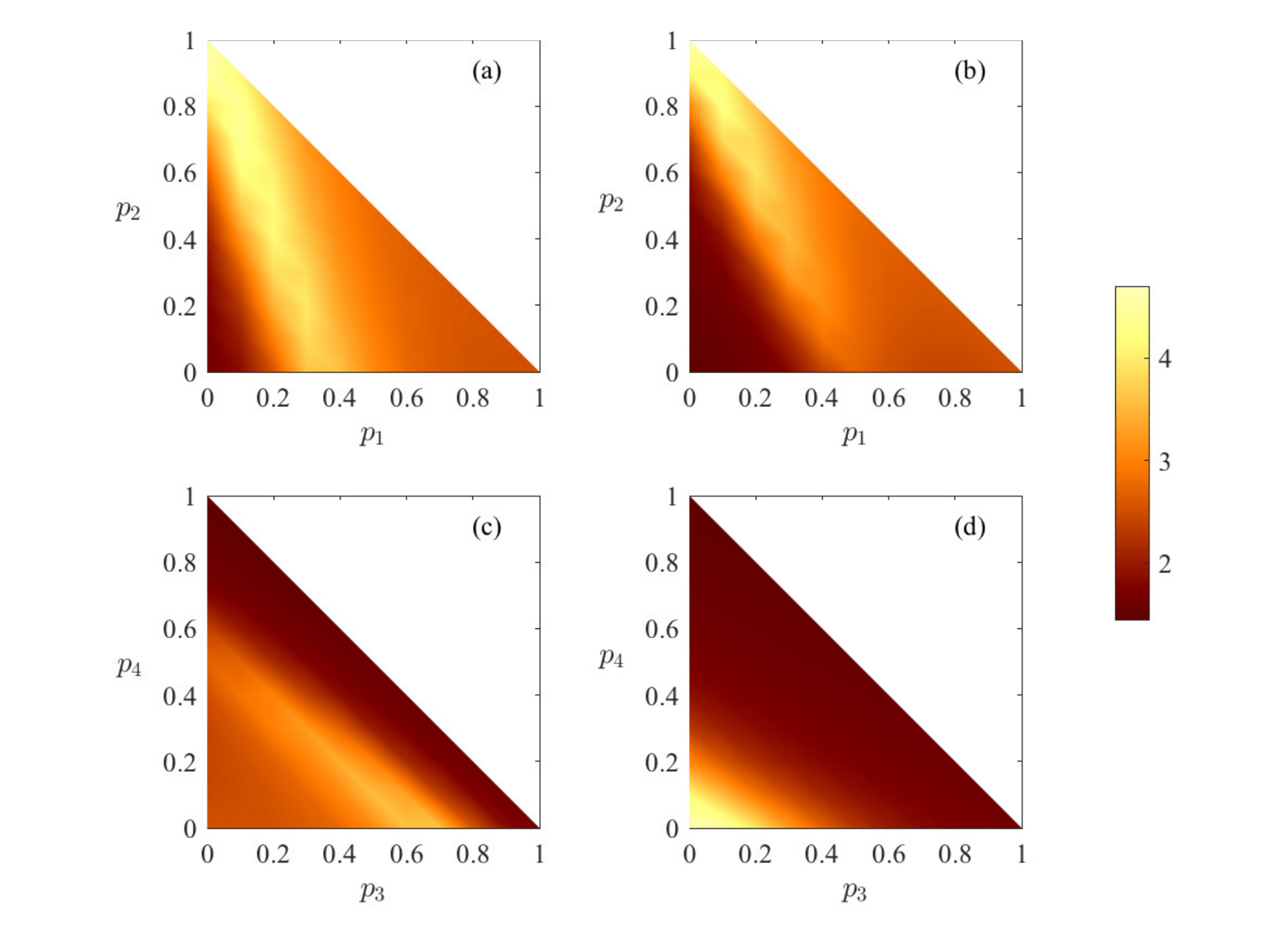}
\caption{ 
Heat maps of $\langle \log_{10} \rho_s \rangle$ for instance classes
composed of mixtures of three subproblem types: (a) $C_1$--$C_2$--$C_3$
mixtures, (b) $C_1$--$C_2$--$C_4$ mixtures, (c) $C_1$--$C_3$--$C_4$
mixtures, and (d) $C_2$--$C_3$--$C_4$ mixtures.  The results are for
system size $L=32$.  Simulations were performed for a finite set of
instance classes defined on a discrete grid with each probability
parameter $p_i$ incremented by a step size of $0.1$. For each of those
instance classes, $\langle \log_{10} \rho_s \rangle$ is obtained by
averaging over the results of $200$ instances and depicted as a color
hue. $\langle \log_{10} \rho_s \rangle$ values for intermediate points
in the parameter space are estimated through interpolation.}
\label{fig:heatmaps}
\end{figure}

Among problems constructed using a single subproblem type, $C_2$ base class instances
are the hardest according to the $\langle \log_{10} \rho_s \rangle$ metric (see 
Fig.~\ref{fig:hist_base_classes} and Table~\ref{tab:hardness_comp}).  Therefore
it is interesting to investigate how problem hardness is affected when mixing $C_2$
plaquettes with other plaquette types.  Fig.~\ref{fig:c2_binary_mixtures} shows
$\langle \log_{10} \rho_s \rangle$ [panel (a)], median SA TTS [panel(b)], and median
SQA TTS [panel (c)] for the three types of binary subproblem mixtures involving
$C_2$: $C_1$--$C_2$, $C_2$--$C_3$, and $C_2$--$C_4$.  The results are for the largest
system $L=32$.  The results are plotted against $p_2$, which is the fraction of $C_2$
plaquettes.  All three hardness metrics indicate that problem hardness monotonically
increases with $p_2$ for all three types of binary mixtures.  Interestingly, one can
observe variations in the relative hardness levels of the three types of binary
mixtures across different $p_2$ values as well as different algorithms.  For example,
SQA finds $C_2$--$C_3$ plaquette mixtures to be harder than both $C_1$--$C_2$ and
$C_2$--$C_4$ mixtures for the entire range of $p_2 \in [0, 1)$, whereas according to
SA TTS and $\langle \log_{10} \rho_s \rangle$ the hardness of $C_1$--$C_2$ mixtures
surpasses that of $C_2$--$C_3$ mixtures as $p_2$ approaches zero.

\begin{table}
\caption{
	A comparison of $\langle \log_{10} \rho_s \rangle$, SA median TTS, 
	and SQA median TTS for a selected set of instance classes.
	The results are for the system size $L=32$.
	\label{tab:hardness_comp}}
\begin{tabular*}{\columnwidth}{@{\extracolsep{\fill}} llrr}
\hline \hline
Instance class & $\langle \log_{10} \rho_s \rangle$ & SA TTS ($\mu$s) & SQA TTS ($\mu$s) \\ [0.5ex] 
\hline
$C_1$           & 2.520(3)   & $4.040(8) \times 10^3$ & $3.11(3) \times 10^6$ \\ 
$C_2$           & 4.70(2)    & $1.9(2) \times 10^7$   & $4.9(6) \times 10^8$ \\
$C_3$           & 1.6447(2)  & $1.45(2) \times 10^3$  & $2.86(6) \times 10^7$\\    
$C_4$           & 1.44822(7) & $2.51(3) \times 10^0$  & $2.00(3) \times 10^5$ \\
\hline
(0.3, 0, 0.7)   & 3.62(2)    & $3.8(3) \times 10^6$   & $6.1(8) \times 10^8$ \\
(0.35, 0, 0.65) & 3.75(3)    & $3.7(2) \times 10^6$   & $4.6(5) \times 10^8$ \\
(0.4, 0, 0.6)   & 3.61(3)    & $2.18(6) \times 10^6$   & $2.8(2) \times 10^8$ \\
\hline
(0.4, 0, 0)     & 2.39(2)    & $2.6(3) \times 10^5$   & $4.4(5) \times 10^8$ \\
(0.45, 0, 0)    & 2.67(2)    & $6.1(7) \times 10^5$   & $4.2(3) \times 10^8$ \\ 
(0.5, 0, 0)     & 2.72(2)    & $5.3(4) \times 10^5$   & $2.8(3) \times 10^8$ \\ [1ex]
\hline \hline
\end{tabular*}
\end{table}

In Table~\ref{tab:hardness_comp}, we present a quantitative comparison of
$\langle \log_{10} \rho_s \rangle$, SA median TTS,
and SQA median TTS for a selected set of instance classes, including the four base classes,
three data points in the vicinity of the hardness peak in $C_1$--$C_3$ mixtures, 
\{(0.3, 0, 0.7), (0.35, 0, 0.65), (0.4, 0, 0.6)\},
and three points close to the hardness peak in $C_1$--$C_4$ mixtures, 
\{(0.4, 0, 0), (0.45, 0, 0), (0.5, 0, 0)\}.
The results are for the system size $L=32$.
When comparing results for the four base classes we find that, 
according to all three algorithms, $C_4$ instances are the easiest to solve
whereas $C_2$ instances are the hardest.
A comparison of the solver performances for the $C_1$ and $C_3$ base classes 
reveals an interesting feature with
regard to the relative hardness levels of the two classes.  According to
the two classical algorithms PAMC and SA, instances from the $C_1$ base
class are harder than those from the $C_3$ class, as conveyed by relatively
higher values of $\langle \log_{10} \rho_s \rangle$ and SA TTS.
However, SQA TTS for the $C_1$ base class is lower than that for $C_3$, 
indicating that SQA finds $C_1$ base class instances easier to
solve than $C_3$ instances.  
We also note that $\langle \log_{10} \rho_s \rangle$ and SA TTS for the $C_2$ class
are noticeably higher than those for the instance classes in the vicinity of
the $C_1$--$C_3$ and $C_1$--$C_4$ transitions.
This indicates that for the system size considered, according to the two 
classical algorithms $C_2$ problems are harder than the instance classes 
near the transitions.
However, when comparing SQA results, we note that the TTS for the $C_2$ class
is within the error bars of the data points with the highest TTS values near 
the $C_1$--$C_3$ and $C_1$--$C_4$ transitions, i.e., (0.3, 0, 0.7) and (0.4, 0, 0).

\subsection{Mixtures of three subproblem types}

Finally, we explore variations in problem hardness among instance
classes constructed using three subproblem types.  Here we discretize
the parameter space into a grid by incrementing each probability
parameter $p_i$ by a step size of $0.1$.  We perform PAMC simulations
for the instance classes defined on the grid points that represent
mixtures of three subproblem types. $\langle \log_{10} \rho_s \rangle$
is computed by averaging over $200$ problems per instance class.
Figure~\ref{fig:heatmaps} shows $\langle \log_{10} \rho_s \rangle$ for
$L=32$ for the four types of three-plaquette combinations:
$C_1$--$C_2$--$C_3$ [Fig.~\ref{fig:heatmaps}(a)], $C_1$--$C_2$--$C_4$
[Fig.~\ref{fig:heatmaps}(b)], $C_1$--$C_3$--$C_4$
[Fig.~\ref{fig:heatmaps}(c)], and $C_2$--$C_3$--$C_4$
[Fig.~\ref{fig:heatmaps}(d)].  The results are represented as heat maps,
with $\langle \log_{10} \rho_s \rangle$ for the points between the grid
points estimated via interpolation. $x$ and $y$ axes in each plot
represent two of the free probability parameters that uniquely define
instance classes in the mixture. The color grade from dark red to light
yellow represents low (red/dark) to high (yellow/light) values of
$\langle \log_{10} \rho_s \rangle$.

\begin{figure}[tb!]
  \includegraphics[width=\linewidth, trim={0cm 0 0 0}, clip]{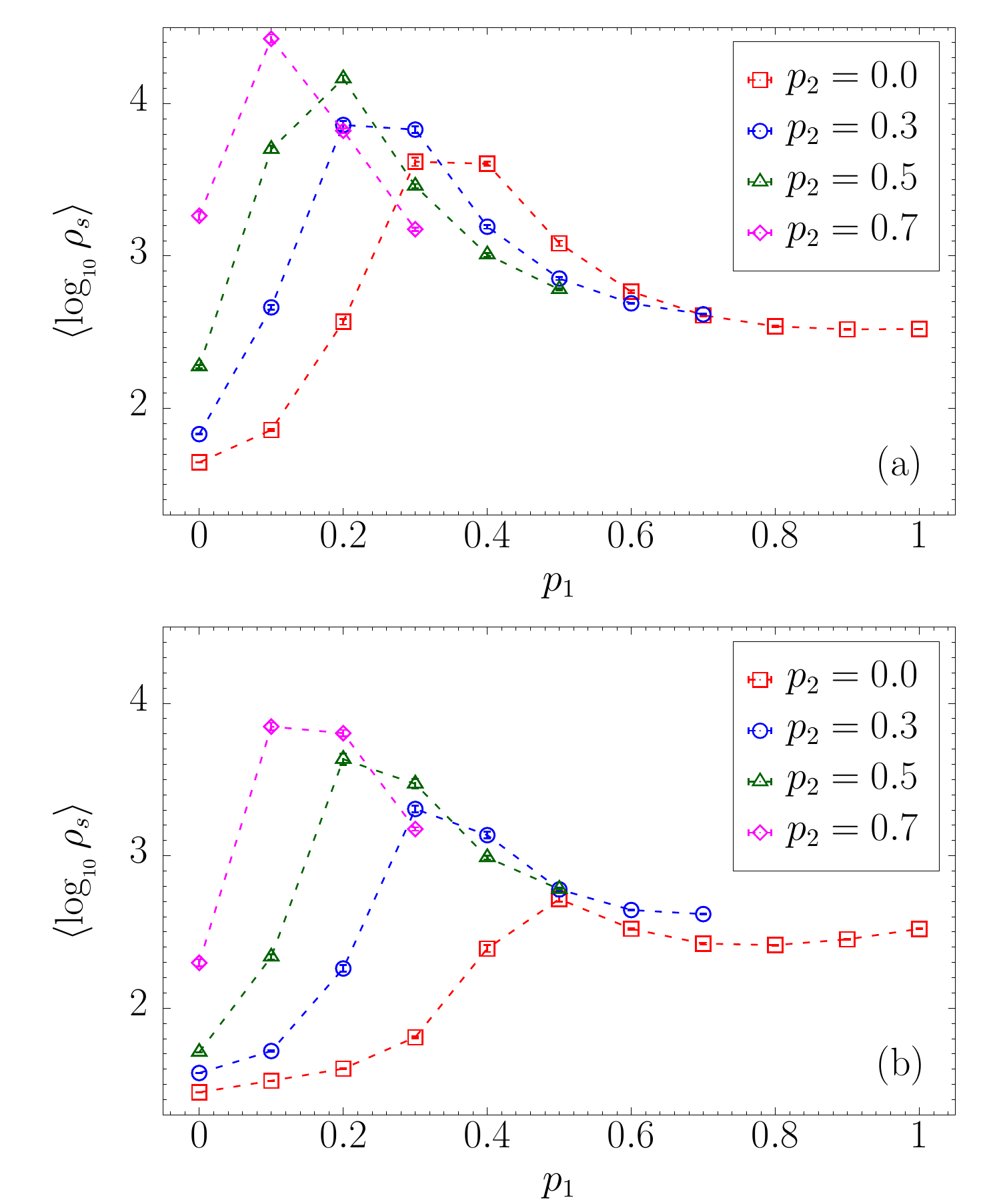}
  \caption{ Population annealing $\langle \log_{10} \rho_s \rangle$ as a 
  function of $p_1$ at fixed values of $p_2$ for (a) $C_1$--$C_2$--$C_3$ mixtures and 
  (b) $C_1$--$C_2$--$C_4$ mixtures. The results are for the system size $L=32$.
   $\langle \log_{10} \rho_s \rangle$ values are estimated using $200$ problems 
   per instance class, per system size. }
 \label{fig:proj_rho_curves_fixed_p2}
\end{figure}

\begin{figure}[tb!]
  \includegraphics[width=\linewidth, trim={0cm 0 0 0}, clip]{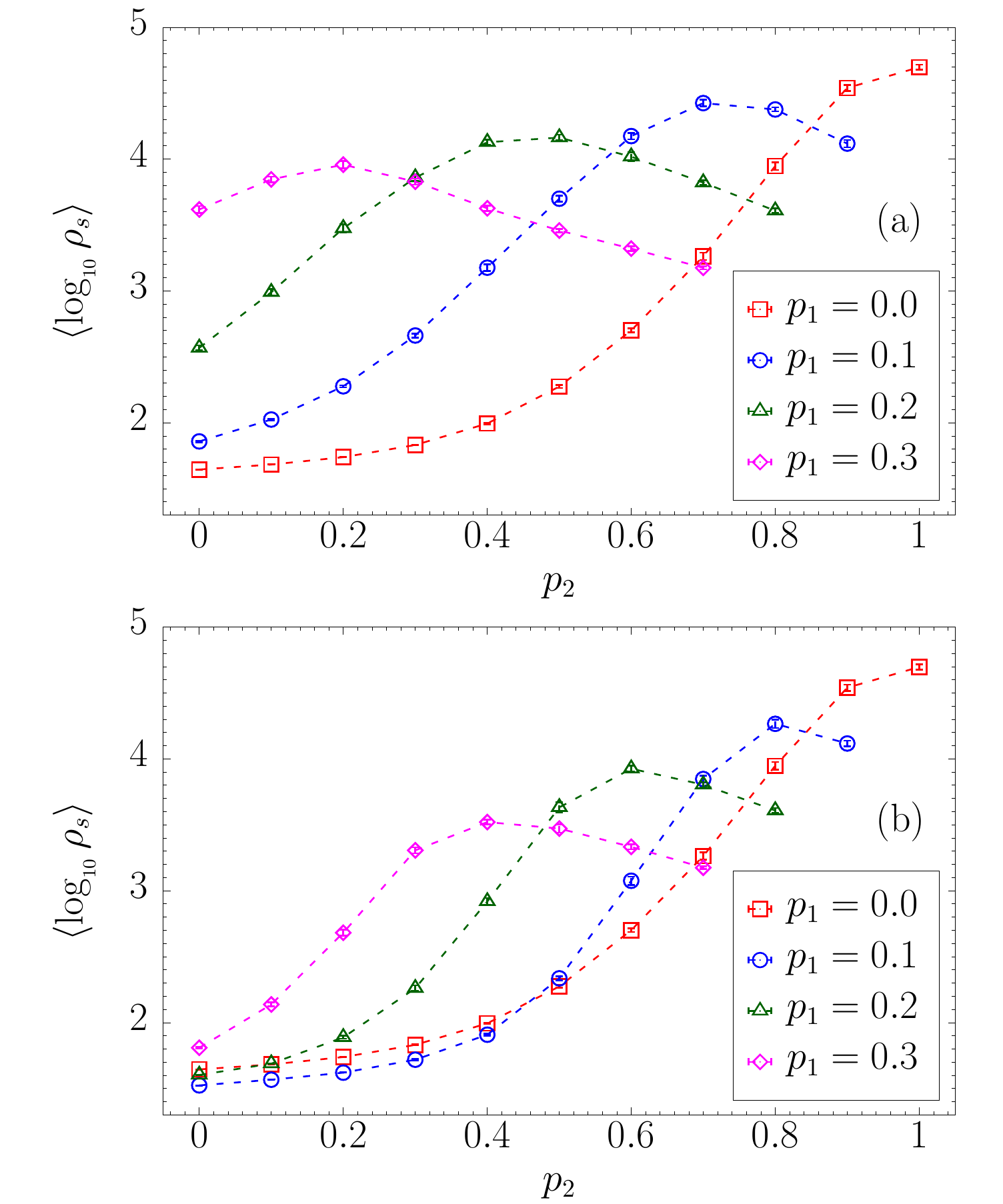}
  \caption{ Population annealing $\langle \log_{10} \rho_s \rangle$ as a 
  function of $p_2$ at fixed values of $p_1$ for (a) $C_1$--$C_2$--$C_3$ mixtures and 
  (b) $C_1$--$C_2$--$C_4$ mixtures. The results are for the system size $L=32$.
   $\langle \log_{10} \rho_s \rangle$ values are estimated using $200$ problems 
   per instance class, per system size. }
 \label{fig:proj_rho_curves_fixed_p1}
\end{figure}

First focusing on $C_1$--$C_2$--$C_3$ [Fig.~\ref{fig:heatmaps}(a)] and
$C_1$--$C_2$--$C_4$ [Fig.~\ref{fig:heatmaps}(b)] mixtures, we note that
$p_2=0$ in Fig.~\ref{fig:heatmaps}(a) and Fig.~\ref{fig:heatmaps}(b),
respectively, correspond to $C_1$--$C_3$ and $C_1$--$C_4$ plaquette
mixtures investigated in Sec.~\ref{sec:two_classes}. For any fixed value
of $p_2$, as $p_1$ approaches from $0$ to $1$, one observes a change in
the color grade of the form dark $\rightarrow$ light $\rightarrow$ dark,
which corresponds to an easy-hard-easy transition.  
For further clarification, we show in Fig.~\ref{fig:proj_rho_curves_fixed_p2} 
the measured values of $\langle \log_{10} \rho_s \rangle$ against $p_1$ 
at fixed $p_2$ values for $C_1$--$C_2$--$C_3$ [panel (a)] and 
$C_1$--$C_2$--$C_4$ [panel (b)] mixtures.
The results clearly show that the hardness
transitions discussed in Sec.~\ref{sec:two_classes} for the case of
$p_2=0$ are also prevalent for nonzero $p_2$ in the multidimensional
parameter space of three-plaquette combinations. 
Generalizing our conclusions on the $C_1$--$C_3$ and $C_1$--$C_4$ phase transitions,
we claim that all easy-hard-easy transitions in the three-dimensional parameter 
space are driven by magnetic ordering transitions.
As discussed in Sec.~\ref{sec:results} B, the number of strong-ferromagnetic (+2) bonds
in each plaquette controls the level of frustration, and the plaquettes progressively
become less frustrated and more ferromagnetic in the order 
$C_4 \rightarrow C_3 \rightarrow C_2 \rightarrow C_1$.
As such, regions in the parameter space with a high concentration of $C_1$ plaquettes 
are predominantly ferromagnetic, whereas regions with a high concentration of $C_4$ and/or $C_3$ 
plaquettes are predominantly disordered.
The light-colored (hard) regions in the heat maps separating the ferromagnetic and disordered problem spaces 
are characterized by either a high concentration of moderately ferromagnetic $C_2$ plaquettes 
or a balanced mixture of $C_1$ and $C_3$/$C_4$ plaquettes which results in 
an overall moderate concentration of +2 bonds.
A transition of the form easy $\rightarrow$ hard $\rightarrow$ easy occurs as one varies 
the subproblem composition such that the overall concentration of +2 bonds changes from
low (disordered) to high (ferromagnetic).

Fig.~\ref{fig:proj_rho_curves_fixed_p2} further shows that 
for both $C_1$--$C_2$--$C_3$ and $C_1$--$C_2$--$C_4$ mixtures
the peak in the hardness shifts to lower $p_1$ values as $p_2$ is increased.
As the fraction of $C_2$ plaquettes increases, the concentration of +2 bonds also increases,
which causes the ferromagnetic transition to occur at lower values of $p_1$.
Consequently, for a fixed value of $p_1$, increasing the fraction of $C_2$ plaquettes 
leads to a transition from a disordered phase to a ferromagnetic phase,
given that the fraction of $C_1$ plaquettes is small enough
so that the initial phase is disordered.
This manifests as an easy-hard-easy transition at fixed $p_1$ values
as can be observed in Fig.~\ref{fig:heatmaps}(a) and (b).
For clarity, in Fig.~\ref{fig:proj_rho_curves_fixed_p1} we show projected curves 
of $\langle \log_{10} \rho_s \rangle$ at fixed $p_1$ values,
in which we can observe the aforementioned transitions as $p_2$ is varied. 
 
Another hardness transition can be observed in the parameter space of
$C_1$--$C_3$--$C_4$ plaquette mixtures [Fig.~\ref{fig:heatmaps}(c)], as either $p_3$
or $p_4$ approaches from $0$ to $1$ (or, equivalently, when decreasing $p_1$ from $1$
to $0$).  This is a result of the decreasing concentration of $C_1$ plaquettes, which
transforms the problem characteristics from predominantly ferromagnetic to
disordered.  $\langle \log_{10} \rho_s \rangle$ for $C_2$--$C_3$--$C_4$ mixtures
[Fig.~\ref{fig:heatmaps}(d)], however, does not show evidence for a easy-hard-easy
transition in the corresponding parameter space.  This is expected since the
parameter space of $C_2$--$C_3$--$C_4$ mixtures does not contain a region with
predominantly ferromagnetic properties due to the complete lack of $C_1$ plaquettes.
The hardness monotonically increases as either $p_3$ or $p_4$ is decreased or, in
other words, as the fraction of $C_2$ plaquettes is increased.

Based on the results of all four types of three-plaquette mixtures, we
identify that for the system sizes considered $\langle \log_{10} \rho_s
\rangle$ for all instance classes are bounded by those of $C_4$ (lowest
value) and $C_2$ (highest value) base classes.  As a final note, we
point out that the quantitative comparison of hardness levels across
different instance classes as characterized by $\langle \log_{10} \rho_s
\rangle$ may not exactly carry over to other solvers. Exploring the
entire parameter space with other solvers using a TTS based metric is
not practical due to the sheer number of instance classes involved, and
thus was not attempted. However, we believe that the general trends in
hardness we observe through $\langle \log_{10} \rho_s \rangle$, such as
the hardness transitions, are common among the solvers considered here
and, more general, quantum-inspired optimization techniques. This is
corroborated by the comparison of multiple solver performances presented
in Sec.~\ref{sec:two_classes} for two sub-regions of the parameter
space, where all three algorithms confirm the presence of hardness
transitions.

\begin{figure}[h] 
\includegraphics[width=\columnwidth, trim={0cm 0 0 0}, clip]{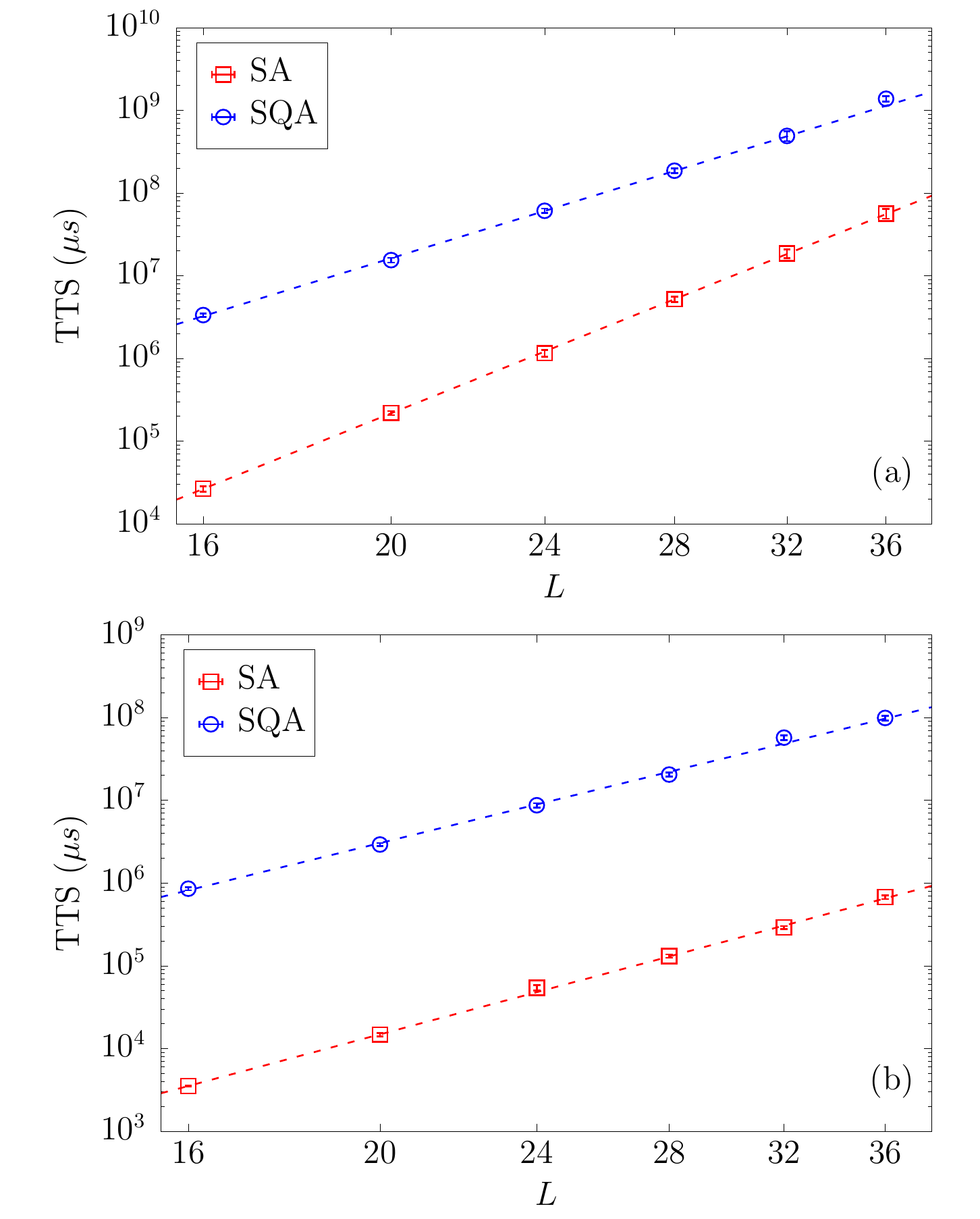}
\caption{ 
Scaling of the median TTS with system size $L$ for simulated 
annealing and simulated quantum annealing for the $C_2$ base class 
[panel (a)] and $C_1$--$C_3$ binary mixtures with $p_1=0.5$ [panel (b)].
Note the logarithmic scale on both axes.
The dashed lines are fits to $\text{TTS} = a L^b$. The scaling coefficients 
for the $C_2$ base class are $b=9.43(4)$ and $b=7.3(1)$ for SA and SQA, 
respectively. For the $C_1$--$C_3$ mixtures, $b=6.44(5)$ for SA and $b=5.9(1)$ 
for SQA. 
}
\label{fig:scaling}
\end{figure}

Given the planarity of the underlying graph, it is clear from general results that
specially crafted algorithms can find the ground state as well as the partition
function in polynomial time \cite{barahona:82}. This does not imply, however, that
more general techniques such as Monte Carlo simulations that are based on (more or
less) local explorations of the energy landscape will show polynomial
scaling. Instead, due to the complex structure of states with many local minima
separated by energy barriers one expects an asymptotically exponential scaling of the
computational effort of such approaches on the systems considered
here. Fig.~\ref{fig:scaling} shows the scaling of the optimal median TTS as a
function of system size $L$ for SA and SQA for two representative instance classes:
the $C_2$ base class [panel (a)] and $C_1$--$C_3$ binary mixtures with $p_1=0.5$
[panel (b)]. We find that our results are best described by polynomial fits of the
form $\text{TTS} = a L^b$. This suggests that for the range of system sizes
considered the scaling of the TTS has not yet reached its asymptotically exponential
behavior.  For the $C_2$ base class, the fits yield scaling coefficients $b=9.43(4)$
for SA, and $b=7.3(1)$ for SQA. For the $C_1$--$C_3$ mixtures, the scaling
coefficients are $b=6.44(5)$ for SA, and $b=5.9(1)$ for SQA. A comparison of the
scaling coefficients shows that for the two instance classes considered
quantum-inspired SQA outperforms classical SA in terms of scaling. However, one should
bear in mind that there are superior classical algorithms for planar graphs, for
example, parallel tempering with isoenergetic cluster moves
(PT+ICM)~\cite{zhu:15b}. In fact, a scaling analysis using PT+ICM for the
$C_1$--$C_3$ mixtures with $p_1=0.5$ has shown that PT+ICM scales better than
SQA~\cite{mandra:unpublished}.

\section{Summary} \label{sec:summary}

We have investigated the computational hardness of planted spin-glass
problems on a square lattice topology generated by the method of ``tile
planting.''  Our planting scheme is based on partitioning the underlying
problem graph into edge-disjoint subgraphs, and embedding subproblems
chosen from predefined classes over the subgraphs. 
Using the PAMC method, we have mapped out a large
region of the hardness phase space. 
For a selected subset of problem classes, we have also performed
SA and SQA calculations
and measured the time-to-solution performance metric.
 
As the building blocks for problem construction, 
we have considered four subproblem types (i.e. frustrated plaquettes)
with different levels of frustration: $C_1$, $C_2$, $C_3$ and $C_4$.
When comparing instances constructed solely using a single subproblem type,
we find instances belonging to the $C_4$ base class to be the easiest,
while we find $C_2$ base class instances to be the hardest.
$C_1$ and $C_3$ base class instances have moderate levels of hardness.
By investigating thermodynamic properties, we have shown that $C_3$ and $C_4$ 
base classes have properties characteristic of disordered spin systems,
whereas the $C_1$ class has strong ferromagnetic properties.
The $C_2$ base class exhibits properties characteristic of both ferromagnetic and 
disordered systems, as indicated by a nonzero (but relatively low) critical temperature 
and a significant ground-state degeneracy. 
We conjecture that the hardness in $C_2$ problems arises as a combined effect 
of the critical slowing down that occurs at the low critical temperature
and the presence of local minima.
We have shown that by mixing different subproblem types
one can achieve tremendous variations in problem hardness.
In particular, we observe several hardness transitions in the phase space,
which can be identified in Fig.~\ref{fig:heatmaps} as 
changes in the color grade of the form dark $\rightarrow$ light $\rightarrow$ dark.
The origin of these transitions can be attributed to underlying 
ferromagnetic phase transitions that occur as the concentration of 
strong-ferromagnetic (+2) bonds is varied.

Due to the
highly tunable hardness, scalability, and ease of implementation, we
believe that our method could be very useful for generating benchmark
problems for novel optimization methods implemented both in hardware and
software.
For the reader interested in constructing benchmark problems, 
we point out that Fig.~\ref{fig:heatmaps} can be used as a visual guide 
for selecting easy versus hard problems, in which easy and hard regimes 
can be identified as dark and light regions, respectively.
In addition, Table~\ref{tab:hardness_comp} presents 
a quantitative comparison of the different performance metrics
for the four base classes and data points close to the hardness transitions
in $C_1$--$C_3$ and $C_1$--$C_4$ binary mixtures.
While $C_2$ instances are the hardest according to the classical algorithms
PAMC and SA, according to SQA results, the instance classes in the vicinity
of the hardness peaks can be as hard as $C_2$ instances within the error bars.
A convenient way of systematically tuning hardness over a wide range is to consider
binary subproblem mixtures involving $C_2$ (see Fig.~\ref{fig:c2_binary_mixtures}),
for which we expect the hardness to increase monotonically as the fraction of $C_2$ 
plaquettes is increased. 
$C_2$--$C_4$ mixtures can be particularly useful in this regard,
as one can tune the hardness from the easiest regime $C_4$ to 
the hardest regime $C_2$, which corresponds to an increase in the time to solution 
over several orders of magnitude in the respective scales for both SA and SQA.

\begin{acknowledgments}

  We thank Mario K\"{o}nz for providing his multispin simulated
  quantum annealing code to perform the calculations presented in this
  paper. D.P.~would like to acknowledge Amin Barzegar and Chris
  Pattison for stimulating discussions.  J.R.~thanks Ryoji Miyazaki
  for informative discussions. F.H.~thanks Cathy McGeogh, Andrew King,
  Paul Bunyk, and Fiona Hanington for helpful remarks. 
  This research is
  based upon work supported in part by the Office of the Director of
  National Intelligence (ODNI), Intelligence Advanced Research
  Projects Activity (IARPA), via MIT Lincoln Laboratory Air Force
  Contract No.~FA8721-05-C-0002.  The views and conclusions contained
  herein are those of the authors and should not be interpreted as
  necessarily representing the official policies or endorsements,
  either expressed or implied, of ODNI, IARPA, or the U.S. Government.
  The U.S. Government is authorized to reproduce and distribute
  reprints for Governmental purpose notwithstanding any copyright
  annotation thereon.  We thank Texas A\&M University and the
  Texas Advanced Computing Center at University of Texas at Austin for
  providing high performance computing resources. M.W.~acknowledges
  support by the European Commission through the IRSES network
  DIONICOS under Contract No.~PIRSES-GA-2013-612707.

\end{acknowledgments}

\bibliography{refs,comments}

\begin{thebibliography}{70}
\bibitem[{com({\natexlab{b}})}]{comment:gauge}
\bibinfo{note}{{We note that any gauge transformations suggested above will not
  change the overall concentration of frustrated plaquettes nor hide the fact
  that all plaquettes of one sublattice are frustrated.}}
\bibitem[{com({\natexlab{c}})}]{comment:lb}
\bibinfo{note}{{Note, however, that to the extent that the chosen population
  sizes are found to be too small for some of the hardest samples, up to
  statistical fluctuations the resulting estimates of $\rho_s$ should be
  considered lower bounds to the actual values of $\rho_s$.}}
\end{thebibliography}

\end{document}